\documentstyle[12pt,epsf]{article}
\def\case{\protect\@case}

\def\@case#1#2{%
\def\@tempa{#2}\def\@tempb{/}%
\ifx\@tempa\@tempb %
\def\@tempa{\@@case{#1}}%
\else %
\def\@tempa{\@@case{#1}{#2}}%
\fi
\@tempa
}

\def\@@case#1#2{{\textstyle{#1\over#2}}}

\long\def\capfont{\@setsize\capfont{10pt}\xpt\@xpt}

\long\def\@makecaption#1#2{%
   \vskip 10\p@
   \setbox\@tempboxa\hbox{\capfont#1: #2}%
   \ifdim \wd\@tempboxa >\hsize
   \begin{quote}
     \capfont  #1: #2\par
   \end{quote}
     \else
       \centerline{\hfil\box\@tempboxa\hfil}%
   \fi}

\begin{document}
\title{\bf Baryon Current Matrix Elements in a Light-Front
Framework\thanks{Submitted to {\it Phys Rev D.}}}
\author{Simon Capstick\\[1.5ex]
Continuous Electron Beam Accelerator Facility,\\
12000 Jefferson Ave., Newport News, VA 23606\\
and\\
Supercomputer Computations Research Institute\\
and Department of Physics,\\
Florida~State~University, Tallahassee, FL
32306\thanks{Present address.}\\[3ex]
B.~D.~Keister\\[1.5ex]
Department of Physics,\\
Carnegie Mellon University, Pittsburgh, PA
15213\thanks{Permanent address.}\\
and\\
Physics Division, National Science Foundation,\\
4201 Wilson Blvd., Arlington, VA 22230}
\date{}
\maketitle
\begin{abstract}
Current matrix elements and observables for electro- and
photo-excitation of baryons from the nucleon are studied in a
light-front framework. Relativistic effects are estimated by
comparison to a nonrelativistic model, where we use simple basis
states to represent the baryon wavefunctions. Sizeable relativistic
effects are found for certain transitions, for example, to radial
excitations such as that conventionally used to describe to the Roper
resonance. A systematic study shows that the violation of rotational
covariance of the baryon transition matrix elements stemming from the
use of one-body currents is generally small.
\end{abstract}
%
\vspace{.25in}
\section{Introduction}
Much of what we know about excited baryon states has grown out
of simple nonrelativistic quark models of their structure. These models were
originally proposed to explain the systematics in the photocouplings of these
states, which are extracted by partial-wave analysis of single-pion
photoproduction experiments. This method can only give us information
about baryon states which have already been produced in $\pi N$ elastic
scattering, since a knowledge of the coupling constant for the
outgoing $N\pi$ channel is necessary for the photocoupling extraction.
Photoproduction experiments have also tended to have limited statistics
relative to the $\pi N$ elastic scattering experiments.

The traditional theoretical approach is to describe the
nucleon and its excitations
using wavefunctions from nonrelativistic potential models, which describe
baryons as being made up of `constituent' quarks moving in a confining
potential. This potential is provided by the interquark `glue', which is taken
to be in its adiabatic ground state. The quarks interact at short distance
via one-gluon exchange. The electromagnetic current is also calculated using a
simple nonrelativistic expansion of the single-quark transition
operator, {\it i.e.,}
in the nonrelativistic impulse approximation. Not surprisingly, the resulting
photocouplings and electroexcitation helicity amplitudes are
frame dependent, with the problem becoming more severe when the photon
transfers
more momentum; current continuity is also violated. This is partly due
to the nonrelativistic treatment, and partly due to the lack of two and
three-body currents which must be present in this strongly-bound system.

Much more can be learned about these states from exclusive electroproduction
experiments. The photocouplings are the values of transition form factors
at the real-photon point $Q^2=0$ (here $Q^2=-q^2$, where $q^\mu$
is the four-momentum transferred from the electron);
electroproduction experiments
measure the $Q^2$ dependence of these form factors, and so simultaneously
probe the spatial structure of the excited states and the initial nucleons.
Both
photoproduction and electroproduction experiments can be extended to examine
final states other than $N\pi$, in order to find `missing' states which are
expected in symmetric quark models of baryons but which do not couple
strongly to the $N\pi$ channel~\cite{KI,CR}. Such experiments are currently
being
carried out at lower energies at MIT/Bates and Mainz. Many experiments to
examine these processes up to higher energies and $Q^2$ values will
take place at CEBAF.

The success of nonrelativistic quark models in describing the systematics
of baryon photocouplings does not extend to the electroproduction amplitudes.
The best measured of these amplitudes are those for elastic
electron-nucleon scattering, the nucleon form factors. In a simple
nonrelativistic model the charge radius of the proton is too small by a factor
of almost two, and the form factors fall off too rapidly at larger $Q^2$
values. It has been suggested that the problem with the charge radius is due to
the neglect of relativistic effects~\cite{HISR}. Similarly, the poor
behavior at large $Q^2$ can be improved by going beyond the nonrelativistic
approximation~\cite{ChungCoester,Schlumpf,HJWeberN,Aznaurian} and by
enlarging the limited (Gaussian) basis in which the wavefunctions are
expanded~\cite{CapKarl}. Similar problems exist in the description
of the $Q^2$ dependence of the transition form factors for the
$\Delta{\textstyle{3\over2}}^+(1232)$ resonance, which are also quite
well measured.
Although the experimental information about the transition form factors for
higher mass resonances is limited~\cite{VBurkert}, there are serious
discrepancies between the predictions of the nonrelativistic model and the
extracted amplitudes here also~\cite{CloseLi,WSPR,Caps}.

It is clear that, once the momentum transfer becomes greater than the
mass of the constituent quarks, a relativistic treatment of the
electromagnetic excitation is necessary. A consistent approach
involves two main parts.  First, the three-body relativistic
bound-state problem is solved for the wavefunctions of baryons with
the assumption of three interacting constituent quarks. Then these
wavefunctions are used to calculate the matrix elements of one-, two-
and three-body electromagnetic current operators. Results of model
studies of the relativistic three-nucleon problem are
available~\cite{3Nboundstate}, and the problem of three constituent
quarks in baryons is presently under investigation~\cite{3qboundstate}
(see also the comments in the final section).  A useful first step is
to consider transitions between substates in a general basis in which
such wavefunctions may be expanded, such as a harmonic-oscillator
basis.  One can then examine successive approximations to the baryon
wavefunctions with these un-mixed oscillator basis states as a
starting point.

There are many competing approaches to the calculation of electromagnetic
currents in relativistic bound states; we will use light-front relativistic
Hamiltonian dynamics~\cite{KP} here. The basis of this approach is to treat
the constituents as particles rather than fields. It allows for an
exact relativistic treatment of boosts, by removing the dependence on
the interacting mass operator of the boost generators. As a consequence
certain rotations become interaction dependent. Within this framework, one
can set up a consistent impulse approximation, which minimizes the effects
of many-body currents~\cite{KP}. Current continuity can be enforced, though,
as we shall see, not uniquely, and one can explicitly evaluate the degree
to which the currents have the required rotational covariance
properties~\cite{BKANG}.

Other authors have applied similar models to the study of these problems;
Chung and Coester~\cite{ChungCoester},
Schlumpf~\cite{Schlumpf}, Weber~\cite{HJWeberN}, and
Aznaurian~\cite{Aznaurian} have studied nucleon elastic form factors
in light-cone models with simple nucleon wavefunctions, and these models
have been extended to the electroproduction amplitudes for the
$\Delta(1232)$ resonance by Weber~\cite{HJWeber}, in a model based on
light-cone field theory, and by Aznaurian~\cite{Aznaurian}. Bienkowska,
Dziembowksi and Weber~\cite{BDW} have also applied a light-cone model to the
study of the small electric-quadrupole $E2$ multipole in
$\Delta{\textstyle{3\over2}}^+(1232)$ electroproduction, and
Weber~\cite{HJW1535} has also applied his model to the $S_{11}(1535)$
resonance.

Our goal is to examine the sensitivity of these photoproduction and
electroproduction amplitudes to relativistic effects. Within the crude
approximation of identifying un-mixed harmonic-oscillator wavefunctions
with the states that they represent in zeroth order in the harmonic basis,
we will examine the nucleon form factors, and the photo- and electroproduction
amplitudes for the positive-parity states
$\Delta{\textstyle{3\over2}}^+(1232)$,
the Roper resonance $N{\textstyle{1\over2}}^+(1440)$,
$N{\textstyle{1\over2}}^+(1710)$, $\Delta{\textstyle{3\over2}}^+(1600)$,
and several
negative-parity resonances including $N{\textstyle{1\over2}}^-(1535)$
and $N{\textstyle{3\over2}}^-(1520)$.
We will show that relativistic effects are in general appreciable;
the most striking relativistic effects we have seen are in the
electroproduction amplitudes for radial excitations such as
$N{\textstyle{1\over2}}^+(1440)$ and $\Delta{\textstyle{3\over2}}^+(1600)$.
We also provide a
systematic study of rotational covariance of the baryon transition
matrix elements.

\section{Conventions}
Much of the light-front notation is presented in Ref.~\cite{KP}.  We present
here the salient features needed to describe the calculation of
current matrix elements.

\subsection{Kinematics}
A homogeneous Lorentz transformation $\Lambda$ describes the relation
between between four-vectors, {\it viz.,}
\begin{equation}
p^\mu \to p'{}^\mu = \Lambda^\mu{}_\nu p^\nu.
\label{ADABBA}
\end{equation}
We also make use of the corresponding $SL(2,C)$ representation,
denoted by ${\underline\Lambda}$.  A rotationless boost is given by
\begin{equation}
{\underline L}_c({\bf A}) \equiv \exp[{\textstyle{1\over2}}
({\hat{\bf A}}\cdot{\mbox{\boldmath $\sigma$}})\sinh^{-1}|{\bf A}|].
\label{MDA}
\end{equation}

Light-front kinematic variables are expressed in terms of four-vectors
which are decomposed as $a^\mu=(a^-,{\bf a}_\perp,a^+)$, where
$a^\pm\equiv a^0\pm\hat {\bf n}\cdot{\bf a}$ and $\hat {\bf n}\cdot{\bf
a}_\perp = 0$.  We adopt here the usual convention of setting
$\hat{\bf n}$ parallel to the $z$ axis.  It is also convenient to
define a light-front vector
\begin{equation}
{\tilde{\bf a}} \equiv ({\bf a}_\perp, a^+).
\label{ABB}
\end{equation}
The $SL(2,C)$ representation of a light-front boost is
\begin{equation}
{\underline L}_f({\tilde{\bf A}}) \equiv
\exp({\textstyle{1\over2}}\sigma^3 \ln A^+)
\exp[{\textstyle{1\over2}}(A^1 + i A^2) (\sigma^1 - i\sigma^2)].
\label{MGA}
\end{equation}

\subsection{State Vectors}
Free-particle state vectors
$| {\tilde{\bf p}}\mu\rangle $ are labeled by the light-front vector
${\tilde{\bf p}}$ and
satisfy the mass-shell condition
\begin{equation}
p^-=(m^2+{\bf p}_\perp^2)/p^+.\label{AC}
\end{equation}
They are normalized as follows:
\begin{equation}
\langle {\tilde{\bf p}}' \mu'| {\tilde{\bf p}} \mu\rangle
= (2\pi)^3 \delta_{\mu'\mu} \delta({\tilde{\bf p}}' - {\tilde{\bf p}})
= (2\pi)^3 \delta_{\mu'\mu}
\delta({\bf p}'_\perp - {\bf p}_\perp)\delta(p'{}^+ - p^+).
\label{AD}
\end{equation}
The state vectors introduced above employ light-front spin.
Under a light-front boost, $p'{}^\mu = \Lambda^\mu{}_\nu p^\nu$, the
state vectors undergo the unitary transformation
\begin{equation}
U(\Lambda) | {\tilde{\bf p}} \mu\rangle
= \sqrt{{p'{}^+}\over p^+}| {\tilde{\bf p}}' \mu\rangle ,
\label{ADA}
\end{equation}
with no accompanying Wigner rotation.
These state vectors are related to this with ordinary (or canonical)
spin via the relation
the relation
\begin{equation}
|{\tilde{\bf p}}\mu\rangle  = \sum_{\bar\mu} \sqrt{\omega_m({\bf p}) \over p^+}
|{\bf p} {\bar\mu}\rangle _c D^{({\textstyle{1\over2}})}_{{\bar\mu}\mu}
[{\underline R}_{cf}(p)],
\label{MFA}
\end{equation}
where the Melosh rotation~\cite{Melosh} is
\begin{equation}
{\underline R}_{cf}(p) = {\underline L}_c^{-1}({{\bf p}\over m})
{\underline L}_f({{\bf p}\over m}).
\label{MFAAA}
\end{equation}

\subsection{Three-Body Kinematics}
Consider three free particles with masses $m_1$, $m_2$ and $m_3$.  We
label their respective light-front momenta in an arbitrary frame by
${\tilde{\bf p}}_1$, ${\tilde{\bf p}}_2$ and ${\tilde{\bf p}}_3$.
The total light-front
momentum is
\begin{equation}
{\tilde{\bf P}} = {\tilde{\bf p}}_1 + {\tilde{\bf p}}_2 + {\tilde{\bf p}}_3.
\label{KZAA}
\end{equation}
Let ${\bf k}_1$, ${\bf k}_2$ and ${\bf k}_3$ be the ordinary
three-momenta in a frame where the total momentum is zero:
\begin{equation}
\sum_{i=1}^3 {\bf k}_i = 0.
\label{KAA}
\end{equation}
The two sets of momenta are related as follows:
\begin{eqnarray}
{\bf k}_{i\perp} &&= {\bf p}_{i\perp} - x_i {\bf P}_\perp;
\quad x_i = p_i^+ / P^+ ;\nonumber  \\
k_{i3} &&= {\textstyle{1\over2}}\left[x_i M_0
- {m_i^2 + {\bf k}_{i\perp}^2 \over x_i M_0} \right],
\label{KAB}
\end{eqnarray}
where $M_0$ is the invariant mass of the three-particle system:
\begin{eqnarray}
M_0 &&= \sum_{i=1}^3\omega_{m_i}({\bf k}_i)
= \sqrt{P^+P^- - {\bf P}_\perp^2} \nonumber  \\
&& = \left[\sum_{i=1}^3 {m_i^2 + {\bf k}_{i\perp}^2 \over x_i}
\right]^{\textstyle{1\over2}},
\label{KBA}
\end{eqnarray}
where
\begin{equation}
\omega_m({\bf k}) \equiv \sqrt{m^2 + {\bf k}^2}.
\label{KBB}
\end{equation}
The Jacobian of the transformation
$({\tilde{\bf p}}_1, {\tilde{\bf p}}_2, {\tilde{\bf p}}_3)
\to ({\tilde{\bf P}}, {\bf k}_1, {\bf k}_2)$
is
\begin{equation}
\left|{\partial({\tilde{\bf p}}_1, {\tilde{\bf p}}_2, {\tilde{\bf p}}_3)
\over \partial({\tilde{\bf P}}, {\bf k}_1, {\bf k}_2)}\right| =
{p_1^+ p_2^+ p_3^+ M_0 \over
\omega_{m_1}({\bf k}_1)\omega_{m_2}({\bf k}_2)\omega_{m_3}({\bf k}_3)
P^+}.
\label{KCA}
\end{equation}

\section{Current Matrix Elements}

To calculate the current matrix element between initial and final
baryon states, we first expand in sets of free-particle states:
\begin{eqnarray}
\langle M' j; {\tilde{\bf P}}'\mu' | I^+(0) | M j; {\tilde{\bf P}}\mu\rangle
&&= (2\pi)^{-18}\int d{\tilde{\bf p}}_1'
\int d{\tilde{\bf p}}_2'
\int d{\tilde{\bf p}}_3'
\int d{\tilde{\bf p}}_1
\int d{\tilde{\bf p}}_2
\int d{\tilde{\bf p}}_3
\sum \nonumber  \\
&&\quad
\langle M' j'; {\tilde{\bf P}}'\mu'
| {\tilde{\bf p}}'_1 \mu'_1 {\tilde{\bf p}}'_2 \mu'_2
{\tilde{\bf p}}'_3 \mu'_3\rangle
\nonumber  \\
&&\quad\times
\langle  {\tilde{\bf p}}'_1 \mu'_1 {\tilde{\bf p}}'_2 \mu'_2
{\tilde{\bf p}}'_3 \mu'_3 |
I^+(0)
| {\tilde{\bf p}}_1 \mu_1 {\tilde{\bf p}}_2 \mu_2
{\tilde{\bf p}}_3 \mu_3\rangle
\nonumber  \\
&&\quad\times
\langle  {\tilde{\bf p}}_1 \mu_1 {\tilde{\bf p}}_2 \mu_2
{\tilde{\bf p}}_3 \mu_3|
M j; {\tilde{\bf P}}\mu \rangle .
\label{MZAA}
\end{eqnarray}
Here and henceforth, an empty summation sign denotes an implied sum
over all repeated indices.  We compute only the contributions from
one-body matrix elements:
\begin{eqnarray}
&&
\langle  {\tilde{\bf p}}'_1 \mu'_1 {\tilde{\bf p}}'_2 \mu'_2
{\tilde{\bf p}}'_3 \mu'_3 |
I^+(0)
| {\tilde{\bf p}}_1 \mu_1 {\tilde{\bf p}}_2 \mu_2
{\tilde{\bf p}}_3 \mu_3\rangle  \nonumber  \\
&&\qquad \approx
\sum_{i=1}^3
\langle  {\tilde{\bf p}}_i' \mu_i'| I^+(0) | {\tilde{\bf p}}_i \mu_i\rangle
\prod_{j \ne i}\left[
\delta_{\mu_j'\mu_j}
(2\pi)^3\delta({\tilde{\bf p}}_j' - {\tilde{\bf p}}_j)\right] .
\label{MZAB}
\end{eqnarray}
The matrix elements for the struck quark are then written in terms
of the Pauli and Dirac form factors for the constituent quarks
\begin{equation}
\langle {\tilde{\bf p}}'\mu' | I^+(0) |{\tilde{\bf p}}\mu\rangle
=F_{1q}(Q^2)\delta_{\mu' \mu}-i(\sigma_y)_{\mu' \mu}{Q\over 2m_i}F_{2q}(Q^2).
\label{qlfme}
\end{equation}
Note that this implies that these current matrix elements depend only on
$Q$ (${\bf q}$ is taken to lie along the x-axis) and {\it not}
on the initial and final momentum of the struck quark, as in the
nonrelativistic model.

The baryon state vectors are in turn related to wave functions as
follows:
\begin{eqnarray}
\langle  {\tilde{\bf p}}_1 \mu_1 {\tilde{\bf p}}_2 \mu_2
{\tilde{\bf p}}_3 \mu_3|
M j; {\tilde{\bf P}}\mu \rangle
&&=
\left|{\partial({\tilde{\bf p}}_1, {\tilde{\bf p}}_2, {\tilde{\bf p}}_3)
\over \partial({\tilde{\bf P}}, {\bf k}_1, {\bf k}_2)}\right|
^{-{1\over2}}
\langle {\textstyle{1\over2}} {\bar\mu}_1{\textstyle{1\over2}} {\bar\mu}_2 |
s_{12}\mu_{12}\rangle
\langle s_{12}\mu_{12} {\textstyle{1\over2}}{\bar\mu}_3 | s \mu_s\rangle
\nonumber  \\
&&\quad\times
\langle l_\rho \mu_\rho l_\lambda \mu_\lambda | L \mu_L \rangle
\langle L \mu_L s \mu_s | j \mu\rangle  \nonumber  \\
&&\quad\times
Y_{l_\rho\mu_\rho}({\hat{\bf k}}_\rho)
Y_{l_\lambda\mu_\lambda}({\hat{\bf K}}_\lambda)
\Phi(k_\rho, K_\lambda) \nonumber  \\
&&\quad\times
D^{({\textstyle{1\over2}}){\dag}}_{{\bar\mu}_1\mu_1}[{\underline
R}_{cf}({k}_1)]
D^{({\textstyle{1\over2}}){\dag}}_{{\bar\mu}_2\mu_2}[{\underline
R}_{cf}({k}_2)]
\nonumber  \\
&&\quad\times
D^{({\textstyle{1\over2}}){\dag}}_{{\bar\mu}_3\mu_3}
[{\underline R}_{cf}({k}_3)],
\label{MZAC}
\end{eqnarray}
The quantum numbers of the state vectors correspond to irreducible
representations of the permutation group.  The spins $(s_{12}, s)$ can
have the values $(0, {\textstyle{1\over2}})$, $(1, {\textstyle{1\over2}})$
and $(1, {3\over2})$,
corresponding to quark-spin wavefunctions with mixed symmetry
($\chi^\rho$ and $\chi^\lambda$) and total symmetry ($\chi^S$),
respectively~\cite{IK}.  The momenta
\begin{eqnarray}
{\bf k}_\rho &&\equiv {1\over\sqrt{2}} ({\bf k}_1 - {\bf k}_2); \nonumber  \\
{\bf K}_\lambda &&\equiv {1\over\sqrt{6}} ({\bf k}_1 + {\bf k}_2 - 2{\bf k}_3)
\label{MZAD}
\end{eqnarray}
preserve the appropriate symmetries under various exchanges of
${\bf k}_1$, ${\bf k}_2$ and ${\bf k}_3$.  Note that ${\bf k}_\rho$
and ${\bf K}_\lambda$ as defined above also correspond to the
nonrelativistic three-body Jacobi momenta.  This is a definition of
convenience, rather than a nonrelativistic approximation.  The three
momenta ${\bf k}_i$ are defined with relativistic kinematics, and the
use of ${\bf k}_\rho$ and ${\bf K}_\lambda$ accounts for the fact that
only two of the ${\bf k}_i$ are independent.  It is also convenient
for keeping track of the exchange symmetry of the three quarks.  This
definition allows use of the usual three-quark harmonic-oscillator
wave functions as a basis, but there is nothing nonrelativistic about
this choice.

\section{Multipole Invariants}

 From the current operator $I^\mu(x)$, we define an auxiliary operator
${\hat I}^s_{\mu_s}(x)$, which has explicit components
\begin{eqnarray}
{\hat I}^1_{\pm 1}(x) &&= \mp{1\over\sqrt{2}}(I^1(x) \pm i I^2(x));
\nonumber  \\
{\hat I}^1_{0}(x) &&= I^3(x); \nonumber  \\
{\hat I}^0_{0}(x) &&= I^0(x).
\label{MAA}
\end{eqnarray}

With these definitions, it was shown in Ref.~\cite{KP} that the current
matrix element between states with canonical spins and instant-form
three-momenta could be written as
follows:
\begin{eqnarray}
&&{}_c\langle  M'j'; {\bf P}'\mu' | {\hat I}^s_{\mu_s}(0)
| Mj; {\bf P}\mu\rangle _c \nonumber  \\
&&\quad= \sqrt{M'\over \omega_{M'}({\bf P}')}
\sqrt{{\omega_M({\bf P}_0)}\over{\omega_M({\bf P})}}
(-1)^s {1\over\sqrt{2}}\sum
\langle {\textstyle{1\over2}} \alpha {\textstyle{1\over2}}
{\dot\beta} | s \mu_s\rangle
\langle {\textstyle{1\over2}} \zeta {\textstyle{1\over2}}
{\dot\eta} | {\bar s} {\bar\mu}_s\rangle  \nonumber  \\
&&\quad\quad\times
\langle l \mu_l {\bar s} {\bar\mu}_s | {\cal J} \mu_{\cal J}\rangle
\langle j {\bar\mu} {\cal J} \mu_{\cal J} | j' \mu'\rangle \nonumber  \\
&&\quad\quad\times
Y^l_{\mu_l}{}^*({\bf\hat P}_0)
[{\underline L}^{-1}_c({{\bf P}'\over M'})]_{\zeta\alpha}
[{\underline L}_c({{\bf P}'\over M'})]^{\dag}_{{\dot\eta}{\dot\beta}}
D^{j{\dag}}_{{\bar\mu}\mu}[{\underline R}_c
({\underline L}_c({{\bf P}'\over M'}),P_0)]
\langle M'j' \parallel I_{l{\bar s}{\cal J}}(q^2) \parallel Mj\rangle ,
\label{MCA}
\end{eqnarray}
where
\begin{equation}
P_0 = L_c^{-1}({{\bf P}'\over M'}) P.
\label{MEA}
\end{equation}
The Wigner rotation is
\begin{equation}
{\underline R}_c({\underline L}_c({{\bf P}'\over M'}),P_0) \equiv
{\underline L}_c^{-1}({{\bf P}\over M})
{\underline L}_c({{\bf P}'\over M'})
{\underline L}_c({{\bf P}_0\over M})
\label{MEAA}
\end{equation}

All dynamical information is contained in the reduced matrix element
$\langle M'j' \parallel I_{l{\bar s}{\cal J}}(q^2) \parallel Mj\rangle $.  The
number of independent matrix elements is limited by the number of
combinations of $l$, ${\bar s}$ and ${\cal J}$ which can couple
together.  Parity considerations provide the additional restriction:
\begin{equation}
\Pi'\Pi (-1)^{l+{\bar s}} = +1.
\label{MEAB}
\end{equation}
where $\Pi'$ and $\Pi$ are the intrinsic parities of the final and
initial states, respectively.  Time reversal provides another
constraint for the case of elastic scattering.  Finally, the
continuity equation:
\begin{equation}
q_\mu \langle M' P' \mu' | I^\mu(0) | M P \mu\rangle  = 0
\label{MEAC}
\end{equation}
further restricts the number of independent matrix elements.

The matrix element of $I^+(0)$ between light-front state vectors is
\begin{eqnarray}
&&\langle  M'j'; {\bf P}'\mu' | I^+(0)
| Mj; {\bf P}\mu\rangle  \nonumber  \\
&&\quad= \sqrt{M'\over P'{}^+}
\sqrt{{\omega_M({\bf P}_0)}\over P^+}
{1\over\sqrt{2}}\sum
\left[
-\langle {\textstyle{1\over2}} \alpha {\textstyle{1\over2}} {\dot\beta}
|1 0\rangle
+\langle {\textstyle{1\over2}} \alpha {\textstyle{1\over2}} {\dot\beta}
|0 0\rangle
\right]
\langle {\textstyle{1\over2}} \zeta {\textstyle{1\over2}} {\dot\eta}
| {\bar s} {\bar\mu}_s\rangle  \nonumber  \\
&&\quad\quad\times
\langle l \mu_l {\bar s} {\bar\mu}_s | {\cal J} \mu_{\cal J}\rangle
\langle j {\bar\mu} {\cal J} \mu_{\cal J} | j' \mu'\rangle \nonumber  \\
&&\quad\quad\times
Y^l_{\mu_l}{}^*({\bf\hat P}_0)
[{\underline L}^{-1}_f({{\tilde{\bf P}}'\over M'})]_{\zeta\alpha}
[{\underline L}_f({{\tilde{\bf P}}'\over M'})]^{\dag}_{{\dot\eta}{\dot\beta}}
D^{j{\dag}}_{{\bar\mu}\mu}[{\underline R}_{cf}(P_0)]
\langle M'j' \parallel I_{l{\bar s}{\cal J}}(q^2) \parallel Mj\rangle ,
\label{MHA}
\end{eqnarray}

A knowledge of the reduced matrix elements
$\langle M'j' \parallel I_{l{\bar s}{\cal J}}(q^2) \parallel Mj\rangle $ is
sufficient for computing any observable for baryon electroexcitation.
Furthermore, it is sufficient to compute the matrix elements of
$I^+(0)$ in order to obtain the reduced matrix elements.  To see this,
we show that the matrix elements of the remaining components of
$I^\mu(0)$ can be obtained by suitable transformations of $I^+(0)$
matrix elements.  For spacelike momentum transfer, it is always
possible to find a frame in which $q^+ = 0$ and the spatial momentum
transfer ${\bf q}_\perp$ lies along the $x$ axis.  Given matrix elements
$\langle M'j'; {\tilde{\bf P}}'\mu' | I^+(0) | Mj;
{\tilde{\bf P}}\mu\rangle $ in this frame, the
matrix elements of $I^-(0)$ can be obtained by a rotation of $\pi$
about the $x$ axis:
\begin{equation}
\langle M'j'; {\tilde{\bf P}}'\mu' | I^-(0) | Mj; {\tilde{\bf P}}\mu\rangle
= \langle M'j'; {\tilde{\bf P}}'\mu' U^{\dag}[{\underline R}_x(\pi)]
| I^+(0) | U[{\underline R}_x(\pi)] Mj; {\tilde{\bf P}}\mu\rangle ,
\label{MJA}
\end{equation}
and the matrix elements of $I^2(0)$ can be obtained by a rotation of
the matrix element
$I^3 = {\textstyle{1\over2}}(I^+ - I^-)$ by ${\pi\over2}$:
\begin{equation}
\langle M'j'; {\tilde{\bf P}}'\mu' | I^2(0) | Mj; {\tilde{\bf P}}\mu\rangle
= \langle M'j'; {\tilde{\bf P}}'\mu' U^{\dag}[{\underline R}_x({\pi\over2})]
| I^3(0) | U[{\underline R}_x({\pi\over2})] Mj; {\tilde{\bf P}}\mu\rangle ,
\label{MKA}
\end{equation}
The matrix element of $I^1(0)$ is constrained by the continuity
equation:
\begin{equation}
\langle M'j'; {\tilde{\bf P}}'\mu' | I^1(0) | Mj; {\tilde{\bf P}}\mu\rangle
= {q^- \over 2Q} \langle M'j'; {\tilde{\bf P}}'\mu' | I^+(0) | Mj;
{\tilde{\bf P}}\mu\rangle ,
\label{MLA}
\end{equation}
where $Q = \sqrt{Q^2}$.

\subsection{The $Q^2\to 0$ limit}

Our calculation should go smoothly over to the real-photon case when
we calculate in the limit $Q^2\to0$. Note that since $q^+=0$, we have
$Q^2=-{\bf q}_\perp^2$.  This means that the spatial momentum transfer
${\bf q}_\perp$ vanishes as $Q^2 \to 0$, and as a consequence all of the
light-front matrix elements
$\langle M'j'; {\tilde{\bf P}}'\mu' | I^+(0) | Mj; {\tilde{\bf P}}\mu\rangle $
vanish due to orthogonality between the
initial and final states.  In this limit, the continuity equation,
Eq.~(\ref{MLA}), becomes
\begin{equation}
\langle M'j'; {\tilde{\bf P}}'\mu' | I^1(0) | Mj; {\tilde{\bf P}}\mu\rangle
= {q^- \over 2q_\perp} \langle M'j';
{\tilde{\bf P}}'\mu' | I^+(0) | Mj; {\tilde{\bf P}}\mu\rangle ,
\label{photoncont}
\end{equation}
and $q^\mu=(q^0,0,0,-q^0)=(q^-,{\bf q}_\perp={\bf 0},q^+=0)$, so the
spatial component of $q^\mu$ lies along the $z$-axis. This means that
the matrix elements of $I^1(0)$ give the desired transverse-photon
amplitudes. Since both the matrix elements of $I^+(0)$ and
${\bf q}_\perp$ are tending to zero, we can rewrite
Eq.~(\ref{photoncont})
\begin{equation}
\langle M'j'; {\tilde{\bf P}}'\mu' | I^1(0) | Mj; {\tilde{\bf P}}\mu\rangle
= \lim_{Q^2\to 0} {q^-\over 2}
{\partial\over \partial q_\perp}
\langle M'j'; {\tilde{\bf P}}'\mu' | I^+(0) | Mj; {\tilde{\bf P}}\mu\rangle .
\label{photonlim}
\end{equation}

In practice, rather than calculating a derivative, we use the
multipole amplitudes to compute helicity amplitudes, and the latter
have smooth behavior as $Q^2\to 0$.

Note that if current continuity is imposed on our calculation at the
level of the light-front matrix elements, {\it i.e.,} adherence to
Eq.~(\ref{MLA}) is ensured by writing the $I^1(0)$ matrix elements in
terms of those of $I^+(0)$ using Eq.(~\ref{photoncont}), then we will
have a singularity at $Q^2=-q_\perp^2=0$. Implementing
Eq.~(\ref{photonlim}) is technically difficult when the $I^+(0)$ matrix
elements are evaluated approximately. We have therefore chosen
({\it e.g.,} for the $N$ to $\Delta$ transition, which has three independent
multipole amplitudes) to calculate three of the four multipole
invariants from three of the four $I^+(0)$ matrix elements, and we
impose continuity at this level by writing the fourth invariant in
terms of the other three.

In practice we have chosen to eliminate the light-front matrix element
corresponding to the largest change $\vert \mu^\prime-\mu \vert$ in
magnetic quantum number for the light-front state vectors in
Eq.~(\ref{MHA}). The choice of which multipole to constrain is then
usually obvious; for example in the case of a transition between a
nucleon an excited state with $J^P={\textstyle{1\over2}}^+$, there are
three multipoles: the Coulomb multipole with $(l,{\bar s},{\cal J})$=(0,0,0),
magnetic (1,1,1), and electric multipoles (1,1,0). Clearly we cannot
eliminate the magnetic multipole, since continuity relates the Coulomb
and longitudinal matrix elements. Elimination of the Coulomb multipole
in favor of the electric multipole leads to unphysical results for
$C_{\textstyle{1\over2}}$ at intermediate values of $Q^2$ (where $q^0=0$).
Therefore, in this case we constrain the electric multipole by imposition
of current continuity.

Similarly, in the case of $J^P={\textstyle{3\over2}}^+$ states, we have
the magnetic $(l,{\bar s},{\cal J})$=(1,1,1), Coulomb (2,0,2), electric
(1,1,2), and  second electric (3,1,2) multipoles. Once again a physical
solution cannot involve elimination of the magnetic or Coulomb multipoles;
we have chosen to eliminate the electric multipole with highest $l$ by
imposition of continuity. Making the other choice changes our results for
the helicity amplitudes by an amount comparable to or smaller than the
rotational covariance condition described and evaluated in what follows.
Similarly, for $J^P={\textstyle{1\over2}}^-$ and ${\textstyle{3\over2}}^-$
states we also eliminate the electric multipole with highest $l$.

\subsection{Further conditions on the matrix elements}

While the matrix elements of $I^+(0)$ are sufficient to determine the
reduced matrix elements introduced in Eq.~(\ref{MCA}), they are in
fact not independent of each other.  Parity considerations imply that
\begin{equation}
\langle M'j'; {\tilde{\bf P}}' -\mu' | I^+(0) | Mj; {\tilde{\bf P}} -\mu\rangle
= \Pi' \Pi (-1)^{j' - j} (-1)^{\mu' - \mu}
\langle M'j'; {\tilde{\bf P}}'\mu' | I^+(0) | Mj; {\tilde{\bf P}}\mu\rangle
\label{MMA}
\end{equation}
This cuts the number of independent matrix elements in half.  In
addition, there can be constraints which come from the requirement of
rotational covariance of the current operator.  To derive the
constraint conditions, we note that, for matrix elements between
states with canonical spins, in a frame where the three-momenta
${\bf P}'$ and ${\bf P}$ lie along the quantization ($z$) axis,
\begin{equation}
{}_c\langle M'j'; {\bf P}' -\mu' | I^\mu(0) | Mj; {\bf P} -\mu\rangle _c
= 0, \quad |\mu' - \mu| > 1.
\label{MNA}
\end{equation}
That is, helicity must be conserved.  Since Eq.~(\ref{MNA}) must be
satisfied for all components of the current, it must also be satisfied
for matrix elements of $I^+(0)$.  Transforming to light-front momenta
and spins, with momentum transfer along the $x$ axis, we obtain
\begin{equation}
\sum_{\lambda'\lambda}
D^{j{\dag}}_{\mu'\lambda'}({\underline R}'_{ch})
\langle M'j'; {\tilde{\bf P}}'\lambda' | I^+(0) | M j;
{\tilde{\bf P}}\lambda\rangle
D^j_{\lambda\mu}({\underline R}_{ch}) = 0,\quad |\mu'-\mu| \ge 2.
\label{BAA}
\end{equation}
The rotations above are
\begin{equation}
{\underline R}_{ch} = {\underline R}_{cf}({\tilde{\bf P}}, M)
{\underline R}_y({\pi\over2}), \quad
{\underline R}'_{ch} = {\underline R}_{cf}({\tilde{\bf P}}', M)
{\underline R}_y({\pi\over2}),
\label{BAABA}
\end{equation}
where ${\underline R}_{cf}$ is the Melosh rotation of
Eq.~(\ref{MFAAA}) which, together with the rotation
${\underline R}_y({\pi\over2})$, transforms the state vectors from
light-front spin
to helicity.  For elastic scattering, Eq.~(\ref{BAA}) is applicable
only to targets with $j\ge 1$.  For elastic and inelastic scattering
involving higher spins, there is a separate, unique rotational
covariance condition for each pair of helicities whose difference is
two or more.  Thus, for a transition ${1\over2}\to{3\over2}$, there is
a single condition, while for ${1\over2}\to{5\over2}$, there are
three.  Helicity pairs which differ by an overall sign change do not
generate additional conditions.

The requirement of rotational covariance provides a dynamical
constraint which cannot be satisfied without the introduction of
interaction-dependent currents, {\it i.e.,} two-and three-body current
operators.  Thus, while the parity constraints in Eq.~(\ref{MMA}) can
be satisfied in a calculation employing one-body current matrix
elements, the rotational covariance condition in Eq.~(\ref{BAA})
cannot.  A measure of the violation of the condition (and hence the
need for many-body current matrix elements) is then the value of the
left-hand side of Eq.~(\ref{BAA}) for each independent pair of
helicities for which the condition is nontrivial.  In an earlier
work~\cite{BKANG}, it was shown that rotational covariance tends to
break down for constituent models of mesons when ${Q\over 2M} \approx
1$.  For mesons with mass of a few hundred MeV, this limits the
applicability of such a calculation to $Q^2$ less than 1 or 2
GeV${}^2$.  Since baryons are hundreds of MeV heavier than light
mesons, we would expect the violation in this range of $Q^2$ not to be
so severe, but it can be checked directly from the calculated matrix
elements, and is discussed further below.

\section{Results}
The result of combining Eqs.~(\ref{MZAA})--(\ref{MZAC}) is a
six-dimensional integral over two relative three
momenta. These integrations are performed numerically, as the angular
integrations cannot be performed analytically.  The integration
algorithm is the adaptive Monte Carlo method VEGAS~\cite{VEGAS}.
Typical statistical uncertainties are on the order of a few percent
for the largest matrix elements.
In what follows we have
taken point-like constituent quarks, {\it i.e.,} with
$F_{1q}(Q^2)=F_{2q}(Q^2)=1$,
in our evaluation of Eq.(~\ref{qlfme}). The light-quark mass is
taken~\cite{HISR,CI}
to be $m_u=m_d=220$ MeV.

\subsection{Nucleon elastic form factors}
Using the techniques outlined above we can form the light-front current
matrix elements for nucleon elastic scattering
$\langle M_N\, {\textstyle{1\over2}};
{\tilde{\bf P}}'\mu' | I^+(0) | M_N\, {\textstyle{1\over2}};
{\tilde{\bf P}}\mu\rangle $, from
Eq.(~\ref{MZAA}). We have evaluated Eq.(~\ref{MZAC}) using a simple
ground-state harmonic oscillator basis state,
\begin{equation}
\Phi_{0,0}(k_\rho, K_\lambda)={1\over \pi^{\textstyle{3\over2}}\alpha_{HO}^3}
{\rm exp}\left\{-[k_\rho^2+K_\lambda^2]/2\alpha_{HO}^2]\right\},
\label{HOgs}
\end{equation}
where the oscillator size parameter $\alpha_{\rm HO}$ is taken~\cite{IK,KI} to
be $0.41$ GeV. Eq.~(\ref{qlfme}) applies equally
well to quark spinor and nucleon spinor current matrix elements, so we can
extract $F_1(Q^2)$ and $F_2(Q^2)$ for the nucleons directly from the above
light-front matrix elements.

\begin{figure}
\vspace{8cm}
\includegraphics{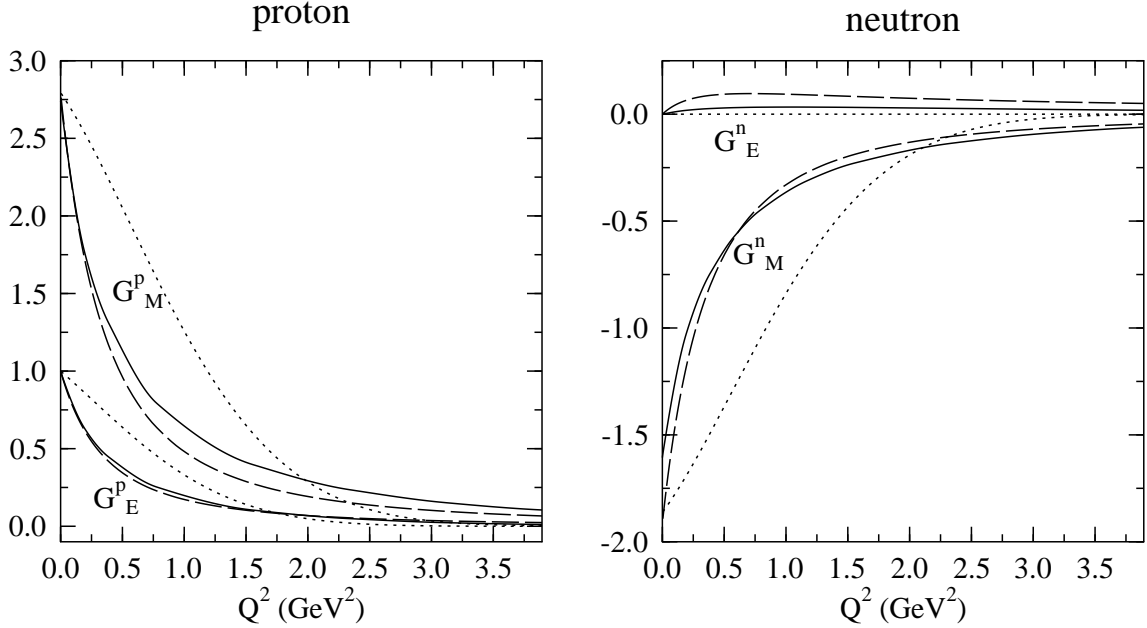}
\caption{Proton and neutron elastic form factors $G_E$ and $G_M$.
The solid curves are the relativistic calculation, the dotted lines give
the corresponding nonrelativistic result, and the dashed lines are the
modified-dipole fit to the data.\label{NGEGM}}
\end{figure}

Figure~\ref{NGEGM} compares the proton and neutron $G_E$ and
$G_M$ calculated in this way, and by using the same wavefunction and the usual
nonrelativistic approach. Also plotted in Fig.~\ref{NGEGM} is
the modified-dipole fit to the data. Our (Breit-frame) nonrelativistic
calculations use a quark mass of $m_{u,d}=336$ MeV (from a
nonrelativistic fit to the nucleon magnetic moments) and the same
oscillator size parameter as above, with the first order nonrelativistic
reduction of the electromagnetic interaction operator
\begin{equation}
H^{\rm nr}=-\sum_{i=1}^3\left\{ {e_i\over 2m}
\left({\bf p}_i\cdot{\bf A}_i+{\bf A}_i\cdot{\bf p}_i\right)
+\mu_i{\mbox{\boldmath $\sigma$}}_i\cdot{\bf B}_i\right\},
\label{Hnr}
\end{equation}
where $m=m_u=m_d$ is the the constituent quark mass,
$e_i$, ${\mbox{\boldmath $\sigma$}}_i/2$, and $\mu_i=ge_i/2m$
are the charge, spin, and magnetic moment of the quark $i$, and
${\bf A}_i\equiv{\bf A}({\bf r}_i)$.

Our choice of
quark mass for the relativistic calculation, while motivated by previous
work~\cite{HISR,CI}, gives a reasonable fit to the nucleon magnetic
moments.
The relativistic calculation yields a proton charge radius close to
that found from the slope near $Q^2$=0 of the dipole fit to the data. The
nonrelativistic calculation falls off too rapidly at larger $Q^2$
[like exp$(-{\bf q}^2/6\alpha_{\rm HO}^2)$], which is not the case for the
relativistic calculation. These observations confirm those made earlier
by Chung and Coester~\cite{ChungCoester} and by Schlumpf~\cite{Schlumpf}.

\subsection{Helicity amplitudes}

For spins other than $j'={\textstyle{1\over2}}$ it is convenient to
compare the results of our calculation to helicity amplitudes. These
are defined in terms of the matrix elements found in Eq.~(\ref{MCA})
from our multipole invariants as follows:
\begin{eqnarray}
A^N_{\textstyle{1\over2}}&=&\zeta\sqrt{4\pi\alpha\over 2K_W}
    \,{_c}\langle M_{N^*}j'; {\bf P}'{\textstyle{1\over2}}\vert
\hat{I}^1_{+1}(0) \vert M_N{\textstyle{1\over2}};
{\bf P}-{\textstyle{1\over2}} \rangle_c\nonumber\\
A^N_{\textstyle{3\over2}}&=&\zeta\sqrt{4\pi\alpha\over 2K_W}
    \,{_c}\langle M_{N^*}j'; {\bf P}'{\textstyle{3\over2}}\vert
\hat{I}^1_{+1}(0) \vert M_N{\textstyle{1\over2}};
{\bf P}{\textstyle{1\over2}} \rangle_c \label{Alambdas}
\\
C^N_{\textstyle{1\over2}}&=&\zeta\sqrt{4\pi\alpha\over 2K_W}
    \,{_c}\langle M_{N^*}j'; {\bf P}'{\textstyle{1\over2}} \vert
\hat{I}^0_0(0) \vert M_N{\textstyle{1\over2}};
{\bf P}{\textstyle{1\over2}} \rangle_c,\nonumber
\end{eqnarray}
where
$\zeta$ is the sign of the $N\pi$ decay amplitude of the resonance $N^*$,
$\alpha=e^2/4\pi \simeq 1/137$, $K_W$ is the equivalent real-photon c.m.
frame three momentum $K_W=(W^2-M_N^2)/2W$, and $\vec{k}_{\rm c.m.}$ is
the virtual-photon c.m. frame three momentum
\begin{equation}
{\bf k}_{\rm c.m.}^2=Q^2+{(W^2-M_N^2-Q^2)^2\over 4W^2}.
\end{equation}
In the above $Q=\sqrt{-q^2}$ is the magnitude of the four-momentum
transfer, and $W$ is the square root of the invariant mass evaluated at
resonance, where $W=M_{N^*}$. Note that the square root factors are
introduced in order that the $Q^2\to 0$ limit of the electroproduction
amplitudes corresponds to the photocoupling amplitude; the only
restriction on these factors is that their limit when $Q^2=0$ is the same
as the normalization factor for an external real photon. The above choices,
therefore, represent a convention.
Other authors calculate the quantity $S_{\textstyle{1\over2}}=\vert
{\bf k}_{\rm c.m.}
\vert  C_{\textstyle{1\over2}} /Q$. Note we use an operator $I^0(0)$
for the 0-component of the electromagnetic current which is defined so
that $\,{_c}\langle M_N {\textstyle{1\over2}};
{\bf P}'{\textstyle{1\over2}} \vert \hat{I}^0_0(0)
\vert M_N{\textstyle{1\over2}}; {\bf P}{\textstyle{1\over2}}
\rangle_c=+1$  at $Q^2=0$, where $N$ is a proton.

\subsubsection{Helicity amplitudes for $\Delta(1232)$ electroexcitation}

\begin{figure}[t]
\vspace{9.0cm}
\includegraphics{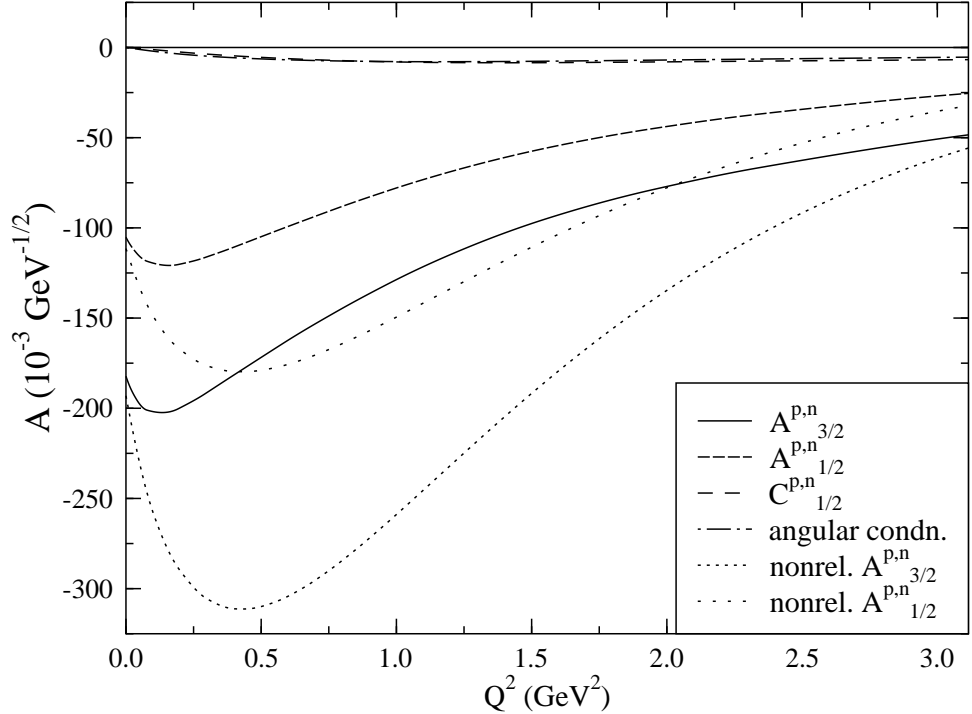}
\caption{$\Delta(1232)$ electroexcitation helicity amplitudes
$A^{p,n}_{3/2}$ (solid curve) and  $A^{p,n}_{1/2}$ (dashed curve),
and $C^{p,n}_{1/2}$ (long dashes). The dotted curves give the corresponding
nonrelativistic results for the Breit-frame transverse helicity amplitudes
($C^{p,n}_{1/2}=0$ in the nonrelativistic model). Also plotted is the value
 of the rotational covariance condition (dot-dashed curve), which happens
to nearly coincide with $C^{p,n}_{1/2}$.\label{Delta}}
\end{figure}

Figure~\ref{Delta} shows our results for the $A_{\textstyle{1\over2}}$,
$A_{\textstyle{3\over2}}$, and $C_{\textstyle{1\over2}}$ helicity
amplitudes for electroexcitation of the $\Delta{\textstyle{3\over2}}^+(1232)$
from nucleon targets, compared to the nonrelativistic results. In this case
we have used the simple single oscillator-basis state from Eq.~(\ref{HOgs})
for both the initial and final momentum-space wavefunctions.
The parameters $\alpha_{HO}$ and $m_{u,d}$ are the same as above.
Although the relativistic calculation does not solve the problem of the
long-standing discrepancy between the measured and predicted photocouplings,
the behavior of the relativistic calculation is closer to the
faster-than-dipole fall off found in the data. The data show no evidence
for the initial rapid rise with $Q^2$ shown by the nonrelativistic
calculation, as pointed out by Foster and Hughes~\cite{FHreview}.

We have also plotted the numerical value of the rotational covariance
condition (multiplied by the normalization factor
$\zeta\sqrt{4\pi\alpha /2K_W}$ for ease of comparison to the physical
amplitudes), given by the left-hand side of Eq.~(\ref{BAA}), for
$\vert \mu^\prime - \mu \vert=2$. At lower values of $Q^2$ the
rotational covariance condition expectation value is a small fraction
of the transverse helicity amplitudes, but approximately the same size
as $C_{\textstyle{1\over2}}$ and larger than the value of $E2/M1$ implied by
our
$A_{\textstyle{1\over2}}$ and $A_{\textstyle{3\over2}}$.  Calculations
which attempt to predict the
ratio using this approach~\cite{BDW} will in general be limited by
rotational covariance uncertainties of similar magnitude.

\subsubsection{Helicity amplitudes for electroexcitation of radially
excited states}

Given the controversy surrounding the nature of the baryon states assigned  to
radial excitations of the nucleon and $\Delta(1232)$ in the nonrelativistic
model~\cite{radexc}, we compare nonrelativistic and relativistic calculations
for simple basis states which can be used to represent the two $P_{11}$
resonances, $N(1440){\textstyle{1\over2}}^+$ and
$N(1710){\textstyle{1\over2}}^+$, as well as the
$\Delta(1600){\textstyle{3\over2}}^+$, which is assigned as a radial excitation
of $\Delta(1232)$ in the nonrelativistic model.

\begin{figure}
\vspace{7.0cm}
\includegraphics{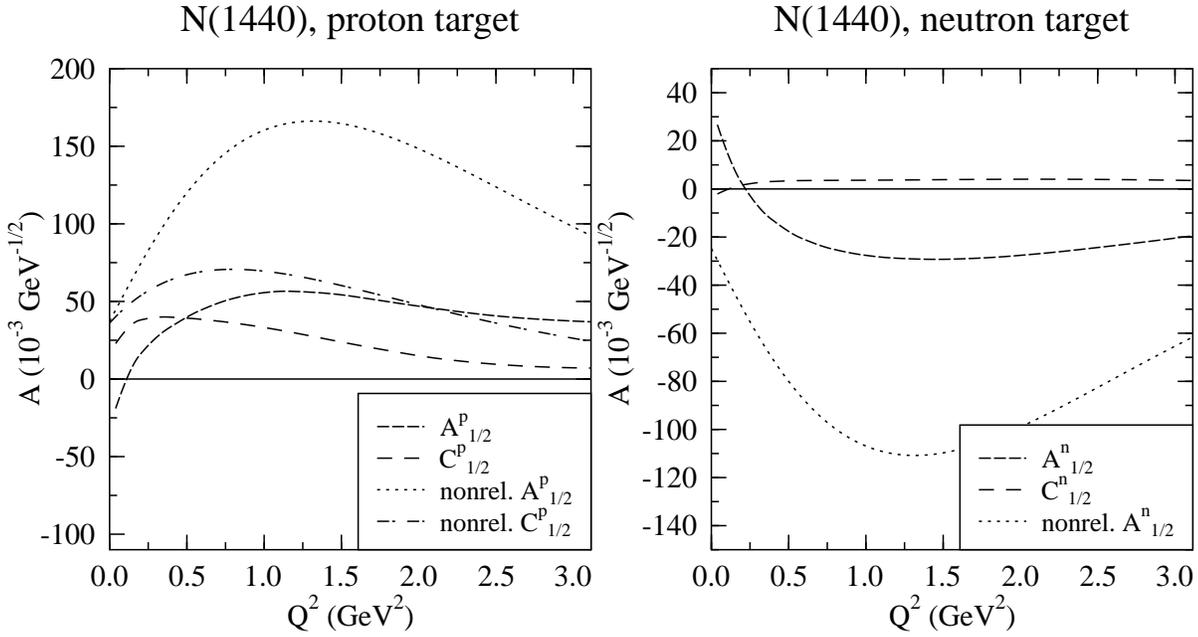}
\caption{$N(1440)$ electroexcitation helicity amplitudes $A^p_{1/2}$
and  $A^n_{1/2}$ (dashed curves), and $C^p_{1/2}$ and $C^n_{1/2}$
(long-dashed curves). The dotted curves give the corresponding
nonrelativistic result for the Breit-frame transverse helicity amplitudes,
and the dot-dashed curve gives the nonrelativistic $C^p_{1/2}$
($C^n_{1/2}$ is zero in the nonrelativistic model).\label{Roper}}
\end{figure}

Figure~\ref{Roper} shows our results for the Roper resonance,
$N{\textstyle{1\over2}}^+(1440)$, for both proton and neutron targets.
Here we have used the single oscillator-basis state Eq.~(\ref{HOgs})
for the initial momentum-space wavefunctions, and a radially excited basis
state
\begin{equation}
\Phi^{S^\prime}_{0,0}(k_\rho, K_\lambda)={1\over \sqrt{3}
\pi^{\textstyle{3\over2}}\alpha_{HO}^3}
\left( 3- [k_\rho^2+K_\lambda^2]/\alpha_{HO}^2 \right)
{\rm exp}\left\{-[k_\rho^2+K_\lambda^2]/2\alpha_{HO}^2]\right\}
\label{HOradial1}
\end{equation}
is used to represent the excited final state. The sign $\zeta$ of the
$N\pi$ decays amplitude used here is calculated in the $^3P_0$ model of
Ref.~\cite{CR}, using exactly the same wavefunctions. As a consequence,
the sign of our nonrelativistic Roper resonance photocouplings
calculation~\cite{Caps} disagrees with the sign which appears in
Ref.~\cite{KI}, where a reduced matrix element was fit to the sign of
the Roper resonance photocouplings.

There are large relativistic effects, with differences between the
relativistic and nonrelativistic calculations of factors of three or
four. Interestingly, the transverse amplitudes also change sign at low
$Q^2$ values approaching the photon point. The large amplitudes at
moderate $Q^2$ predicted by the nonrelativistic model (which are
disfavored by analyses of the available single-pion electroproduction
data~\cite{Stoler}) appear to be an artifact of the nonrelativistic
approximation. This disagreement, and that of the nonrelativistic
photocouplings with those extracted from the data for this
state~\cite{Caps}, have been taken as evidence that the Roper
resonance may not be a simple radial excitation of the quark degrees
of freedom but may contain excited glue~\cite{ZPL,LBL}. The strong
sensitivity to relativistic effects demonstrated here suggests that
this discrepancy for the Roper resonance amplitudes has a number of
possible sources, including relativistic effects.

We also find in the case of proton targets that there is a sizeable
$C^p_{\textstyle{1\over2}}$, reaching a value of about $40\times 10^{-3}$
GeV$^{{\textstyle{1\over2}}}$ at $Q^2$ values between 0.25 and 0.50 GeV$^2$.
Correspondingly, there will be a sizeable longitudinal excitation amplitude.
The nonrelativistic $C^N_{\textstyle{1\over2}}$ amplitudes shown here and in
all other figures are listed (up to the $N\pi$ sign $\zeta$) in Appendix C;
formulae for the nonrelativistic transverse amplitudes (up to the sign $\zeta$)
are tabulated in Ref.~\cite{KI}.

\begin{figure}
\vspace{8.0cm}
\includegraphics{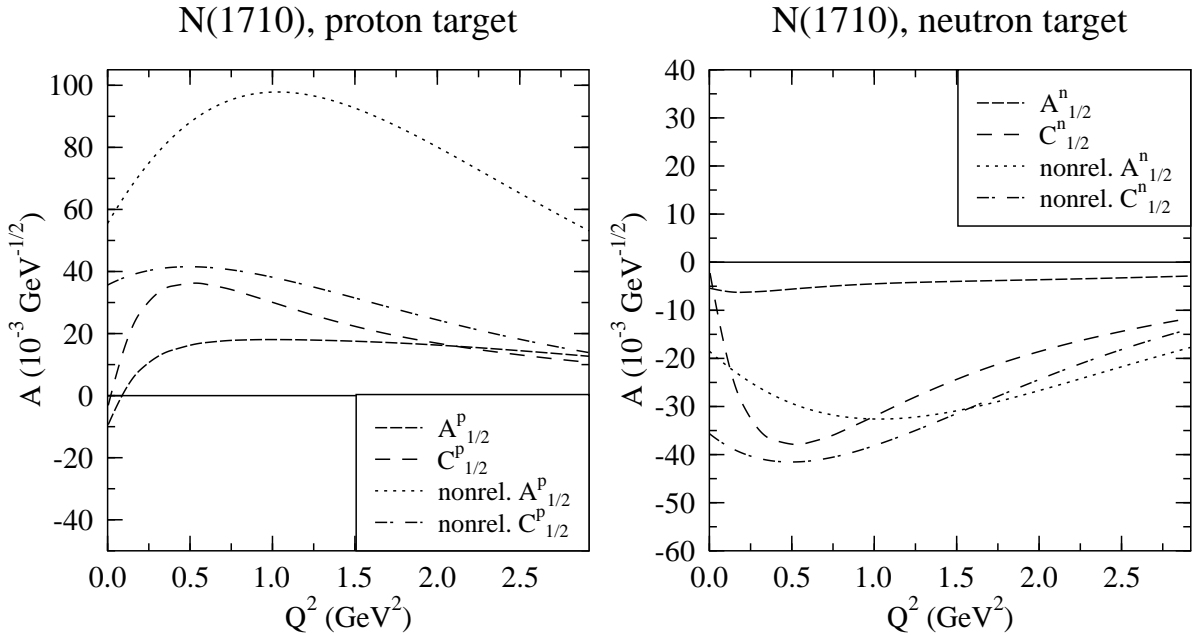}
\caption{$N(1710$ electroexcitation helicity amplitudes. Key as in
Fig.~3, except here the nonrelativistic $C^n_{1/2}$
(dot-dashed curve) is nonzero.\label{N1710}}
\end{figure}

In the case of the $N(1710){\textstyle{1\over2}}^+$, which is assigned a second
radially excited wavefunction in the nonrelativistic model, our results using a
simple wavefunction made up of a linear combination of basis functions (see
Refs.~\cite{IK} and ~\cite{KI}) are shown in Fig.~\ref{N1710}. Here we see an
even greater reduction in the size of the transverse amplitudes in the
relativistic calculation. The $C_{1/2}$ amplitudes at the photon point also are
greatly reduced in size, although at larger $Q^2$ values they become comparable
to the nonrelativistic charge helicity amplitudes, with a similar modification
of the $Q^2$ dependence to that shown above in the case of the Roper resonance
and $G^N_E$.

It is interesting to see whether this pattern is maintained in the case of the
$\Delta(1600){\textstyle{3\over2}}^+$, which has a similar relationship to
$\Delta(1232)$ as the Roper resonance has to the nucleon. This is a quark-spin
${\textstyle{3\over2}}$ state [with spatial wavefunction given by
Eq.~(\ref{HOradial1})] in the nonrelativistic model, and so fundamentally
differs from the $P_{11}$ states described above. However, examination of the
nonrelativistic calculation of the electroexcitation amplitudes shows that in
both cases it is the spin-flip part of the O($p/m$) electromagnetic Hamiltonian
which is responsible for the transverse transition amplitudes.

\begin{figure}
\vspace{9.0cm}
\includegraphics{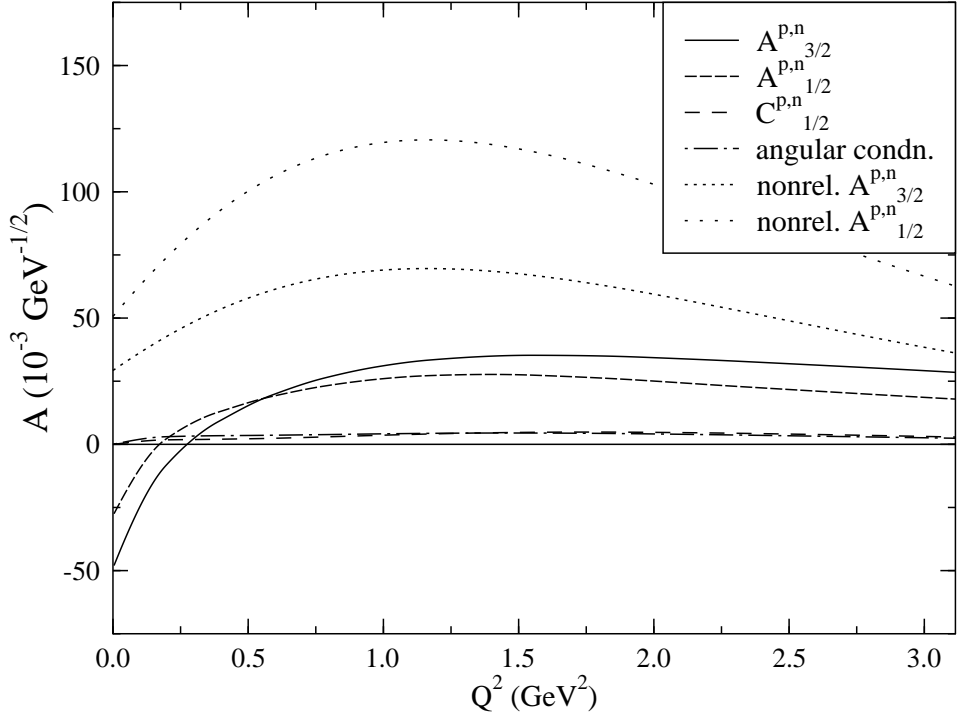}
\caption{$\Delta(1600)$ electroexcitation helicity amplitudes.
Key as in Fig.~2 (here also the nonrelativistic amplitudes $C^{p,n}_{1/2}$
are zero).\label{D1600}}
\end{figure}

This similarity persists in our relativistic calculation (Figure~\ref{D1600}),
where we see photocouplings which have changed sign in comparison to the
nonrelativistic results, along with substantially reduced transverse amplitudes
at intermediate $Q^2$ values. The falloff with $Q^2$ at higher $Q^2$ values is
quite gradual. As in the case of the $\Delta(1232)$, the relativistic $C_{1/2}$
amplitudes (zero in the nonrelativistic model) are small and comparable in size
to the rotational covariance condition, which is generally a small fraction of
the transverse amplitudes.

\subsubsection{Helicity amplitudes for electroexcitation of $P$-wave baryons}

We have also calculated helicity amplitudes for the final states
$N{\textstyle{3\over2}}^-(1520)$ and $N{\textstyle{1\over2}}^-(1535)$,
for both proton and neutron targets. Here we use the same initial-state
wavefunction as above and final state wavefunctions which are made up from
the orbitally-excited momentum-space wavefunctions
\begin{eqnarray}
\Phi^{M_\rho}_{1,+1}&=& -{1\over \pi^{\textstyle{3\over2}}\alpha_{HO}^3}
{k_{\rho +}\over \alpha_{HO}}
{\rm exp}\left\{-[k_\rho^2+K_\lambda^2]/2\alpha_{HO}^2]\right\}\nonumber\\
\Phi^{M_\lambda}_{1,+1}&=& -{1\over \pi^{\textstyle{3\over2}}\alpha_{HO}^3}
{K_{\lambda +}\over \alpha_{HO}}
{\rm exp}\left\{-[k_\rho^2+K_\lambda^2]/2\alpha_{HO}^2]\right\},
\label{HOorbital}
\end{eqnarray}
(where $k_+\equiv k_x+ik_y$) and their counterparts reached by angular
momentum lowering. In this case configuration mixing due to the hyperfine
interaction is included in the final-state wavefunctions. Since these
states are degenerate in mass before the application of spin-spin
interactions, they are substantially mixed by them; details are given in
Appendix B.

\begin{figure}
\vspace{8.0cm}
\includegraphics{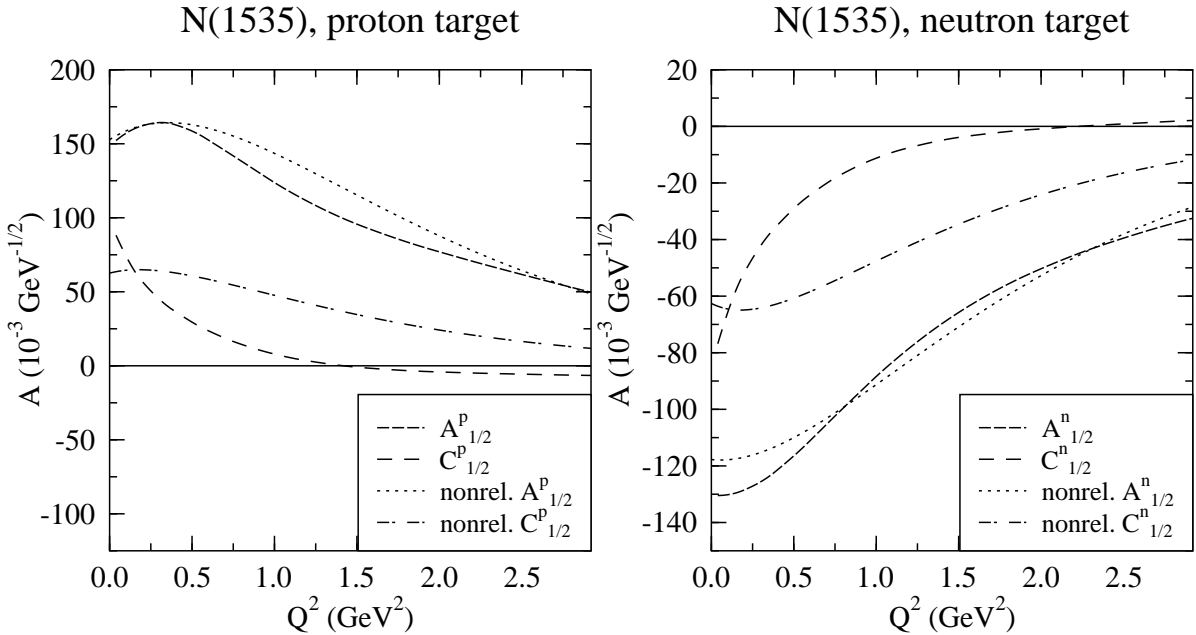}
\caption{$N(1535)$ electroexcitation helicity amplitudes $A^p_{1/2}$ and
$A^n_{1/2}$ (dashed curves), and $C^p_{1/2}$ and $C^n_{1/2}$ (long-dashed
curves). The dotted and dot-dashed curves give the corresponding
nonrelativistic result for the Breit-frame transverse and $C_{1/2}$
helicity amplitudes, respectively.
\label{N1535}}
\end{figure}

The results for the helicity amplitudes for
$N{\textstyle{1\over2}}^-(1535)$ excitation from both proton and
neutron targets are compared to the corresponding nonrelativistic results
in Figure~\ref{N1535}.

In contrast to the results shown above, in this case there appears to
be little sensitivity to relativistic effects in the results for the
transverse amplitudes $A_{1/2}$; this is not the case for the
$C_{1/2}$ amplitudes. For both targets there are substantial $C_{1/2}$
amplitudes at small $Q^2$, and the $Q^2$ dependence is very different in
the relativistic calculation (resembling the dipole behavior of the
nucleon form factors in Fig.~\ref{NGEGM}).

\begin{figure}
\vspace{8.0cm}
\includegraphics{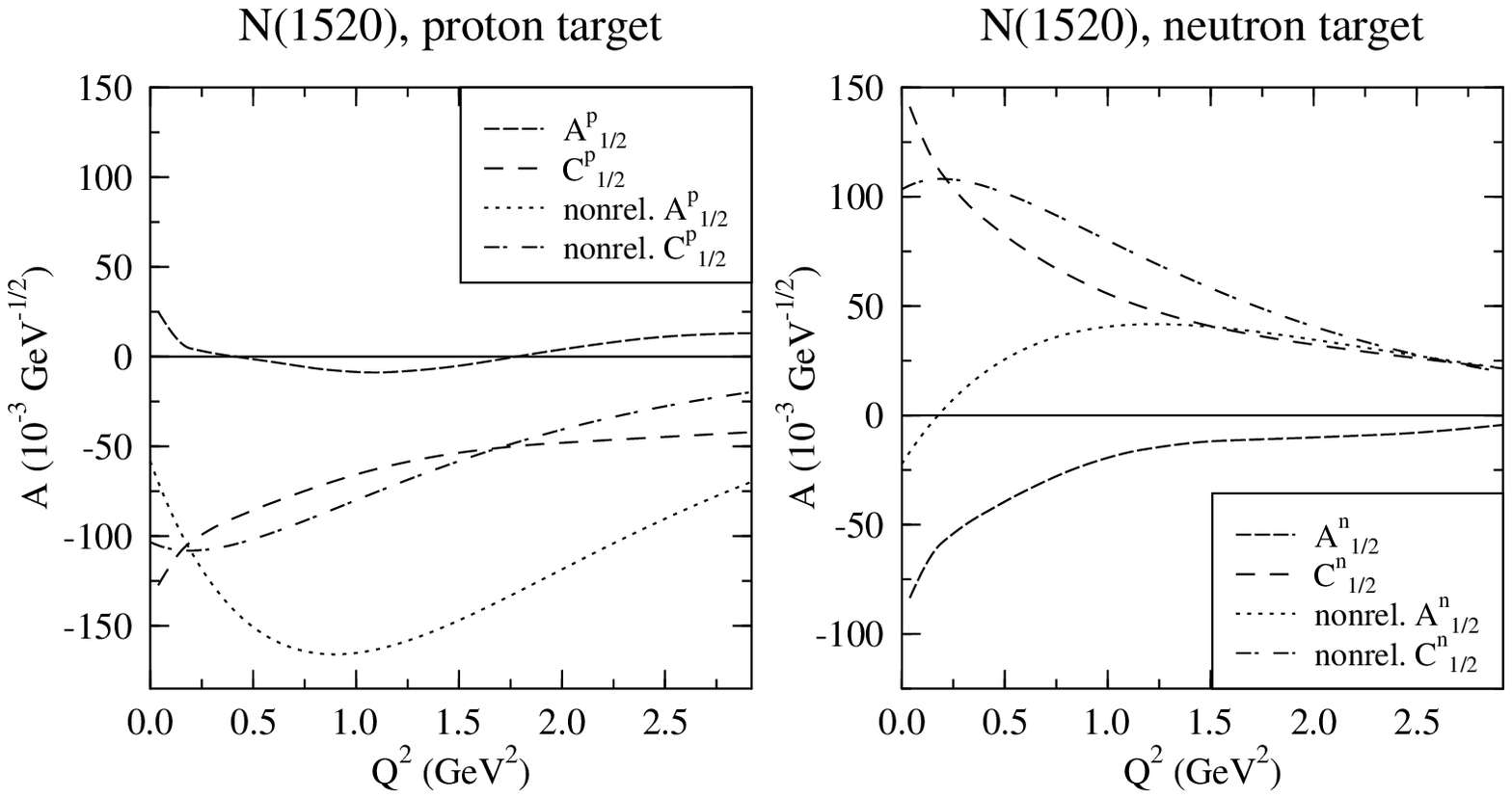}
\caption{$N(1520)$ electroexcitation helicity-${\textstyle{1\over2}}$
amplitudes; key as in Fig.~6.\label{N1520half}}
\end{figure}

\begin{figure}
\vspace{9.0cm}
\includegraphics{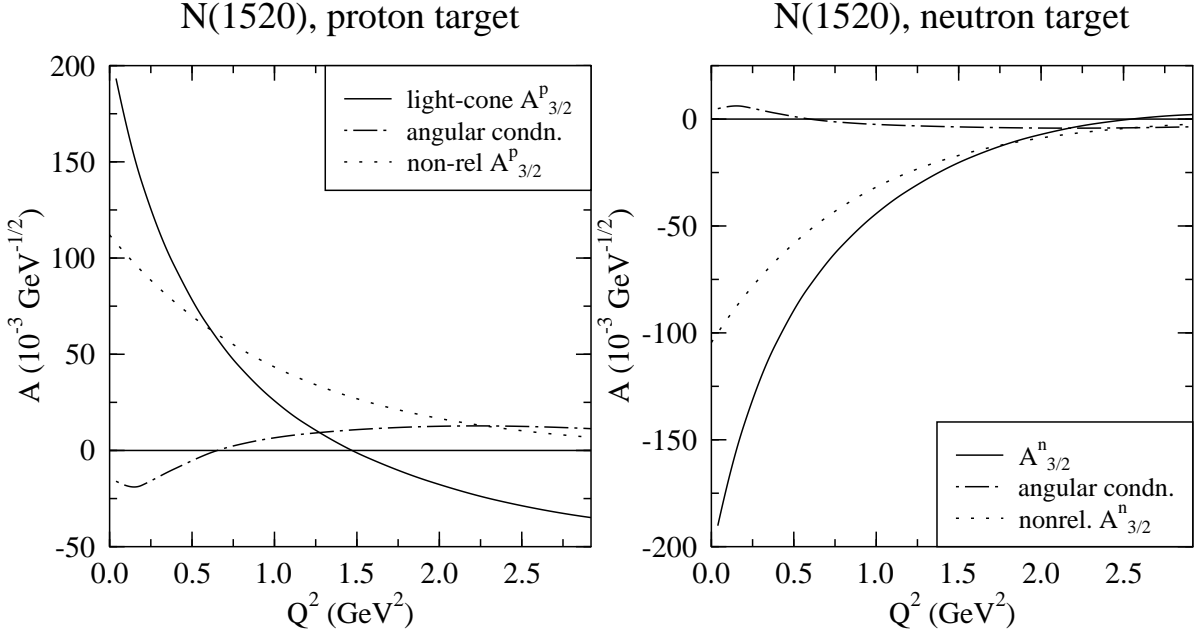}
\caption{$N(1520)$ electroexcitation helicity amplitudes $A^p_{3/2}$ and
$A^n_{3/2}$ (solid curves), and the value of the rotational covariance
condition (dot-dashed curve). The dotted curves give the corresponding
nonrelativistic results for the Breit-frame transverse helicity amplitudes.
\label{N1520thalf}}
\end{figure}

Our results for the $A_{\textstyle{1\over2}}$ and
$C_{\textstyle{1\over2}}$ helicity amplitudes for
$N{\textstyle{3\over2}}^-(1520)$ excitation from both proton and
neutron targets are compared to the nonrelativistic results in
Figure~\ref{N1520half}, and this comparison for $A_{\textstyle{3\over2}}$
and the value of the rotational covariance condition are shown in
Figure~\ref{N1520thalf}.

Here we see large relativistic effects in the $A_{\textstyle{1\over2}}$
amplitudes. From Ref.~\cite{KI} one can see that the
$A_{\textstyle{1\over2}}$ amplitudes for this state are proportional
to $1-k^2/\alpha_{HO}^2$ for a proton target, and  $1-k^2/(3\alpha_{HO}^2)$
for a neutron target, where $k=\vert {\bf k}\vert$ is the virtual-photon
three-momentum. Here the constant term arises from the convection part
of $H_{\rm em}$, and the $k^2/\alpha_{HO}^2$ term arises from the
quark-spin-flip part. Our relativistic treatment can be expected to
change the relationship between these two terms (as well as adding other
effects), and we see here that this has caused a substantial cancellation.
In this case the charge amplitudes $C_{\textstyle{1\over2}}$ are greater
in magnitude than the transverse amplitudes of both proton and neutron
targets, and they are the largest at small $Q^2$.

 From Fig.~\ref{N1520thalf} we can see that the relativistic effects in
$A_{\textstyle{3\over2}}$ are smaller (in the nonrelativistic model
there are not two partially cancelling terms), except near $Q^2=0$.
In the case of a proton target, the rotational covariance condition
expectation value is a substantial fraction of the
$A^p_{\textstyle{3\over2}}$ amplitude at larger values of $Q^2$, and
is comparable to the small $A^p_{\textstyle{1\over2}}$ at all $Q^2$ values;
the absolute size of the rotational covariance condition can be taken to be
a measure of the absolute uncertainty introduced in our results from
considerations of rotational covariance.

\begin{figure}[h]
\vspace{9.0cm}
\includegraphics{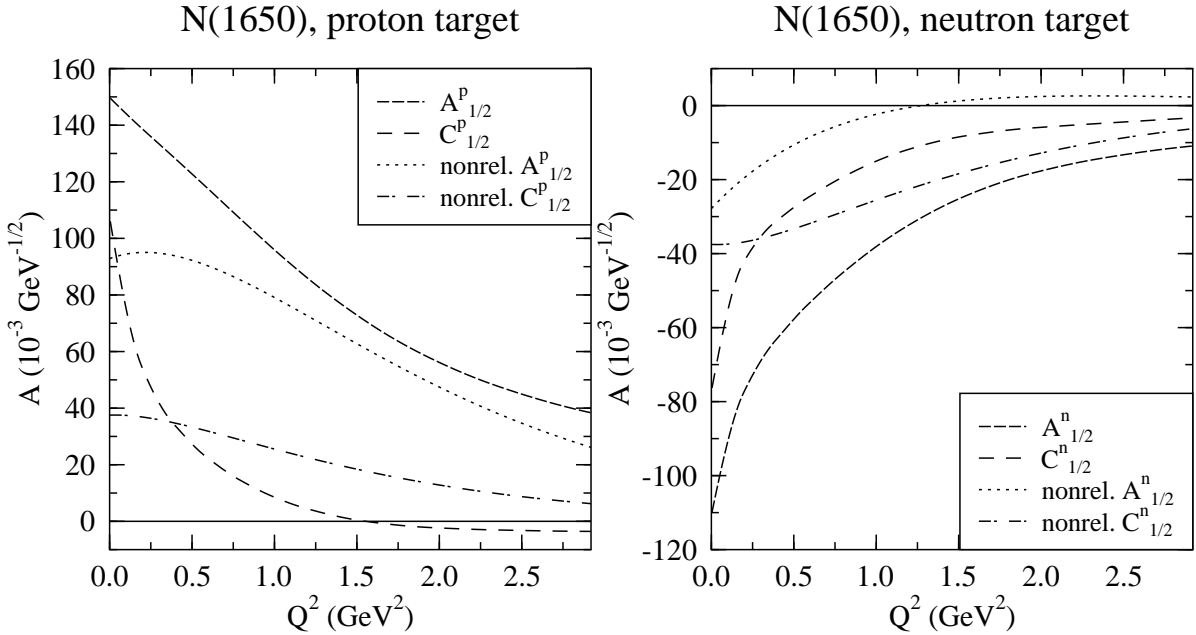}
\caption{$N(1650)$ electroexcitation helicity amplitudes. Key as in
Fig.~6.\label{N1650}}
\end{figure}

As an example of a predominantly quark-spin-${\textstyle{3\over2}}$ P-wave
excitation, Fig.~\ref{N1650} shows relativistic and nonrelativistic amplitudes
for electroexcitation of $N(1650){\textstyle{1\over2}}^-$, the partner of the
predominantly quark-spin-${\textstyle{1\over2}}$ state
$N(1535){\textstyle{1\over2}}^-$ (see Eqs.~(\ref{N4}),~(\ref{N2})
and~(\ref{mixing}) in Appendix B).

Here, in contrast to those of $N(1535){\textstyle{1\over2}}^-$, the transverse
helicity amplitudes show considerable sensitivity to relativistic effects near
$Q^2=0$ for proton targets, and at all $Q^2$ values for neutron targets, and as
above the charge helicity amplitudes have remarkably different $Q^2$ behavior.
Amplitudes for the predominantly quark-spin-${\textstyle{3\over2}}$ state
$N(1700){\textstyle{3\over2}}^-$ tend to be quite small if calculated in the
nonrelativistic model~\cite{{KI},{Caps}}, and this is still true in our
relativistic calculation, with the exception of the amplitude
$A^n_{\textstyle{3\over2}}$.

We complete our survey of relativistic effects for basis functions
representative of $P$-wave states by examining the amplitudes for
electroexcitation of $\Delta(1620){\textstyle{1\over2}}^-$, once again using a
simple linear single-oscillator basis wavefunction [Eq.~(\ref{D2}) in Appendix
B]. Fig.~\ref{D1620} contrasts our results with the nonrelativistic results of
Ref.~\cite{IK} for $A_{\textstyle{1\over2}}$ and for $C_{\textstyle{1\over2}}$
from Appendix C. In both cases the amplitudes near $Q^2=0$ are considerably
larger in the relativistic calculation, with the transverse amplitude
decreasing to below the nonrelativistic result above approximately $Q^2=0.7$
GeV$^2$ due to quite different (dipole-like) behavior as a function of $Q^2$.
Note in the case of negative-parity $\Delta$ states, the charge helicity
amplitudes are not zero if calculated nonrelativistically (see Appendix C). We
have found similar relativistic sensitivity in a calculation using a simple
basis function which can be used to represent the state
$\Delta(1700){\textstyle{3\over2}}^-$.

\begin{figure}
\vspace{9.0cm}
\includegraphics{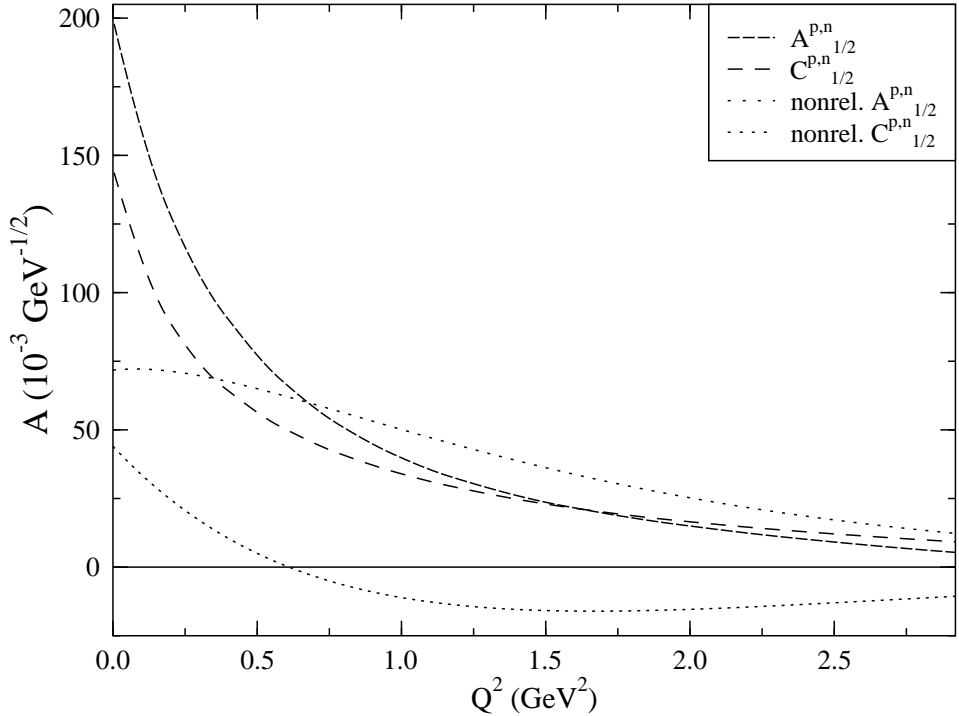}
\caption{$\Delta(1620)$ electroexcitation helicity amplitudes. Key as in
Fig.~6.\label{D1620}}
\end{figure}

\section{Discussion and Summary}

The results outlined above establish that there will be considerable
relativistic effects at all values of $Q^2$ in the electroexcitation amplitudes
of baryon resonances, even at $Q^2=0$. In the real-photon case, this can be
understood in terms of the sizeable photon energy required to photoproduce
these resonances from the nucleon, which implies a photon three-momentum
comparable to the quark Fermi momentum in the nucleon. In particular, our
results show that the $Q^2$ dependence of the nonrelativistic amplitudes is
generally modified into one resembling a dipole falloff behavior, as has been
shown in the case of the nucleon form factors. This behavior is likely to be
partially modified by the inclusion in the wavefunctions of mixings to higher
shells, which are required by any model of baryon structure which takes into
account the anharmonic nature of the confining potential between quarks.
However, we consider it remarkable that relativistic effects account for a
large part of discrepancy between the nonrelativistic model's predictions and
the physical situation.

We wish to stress that the comparisons we have made here, using simple
basis functions representing the individual states, are intended to be
representative of the degree of sensitivity to relativistic effects
for states of their quantum numbers. In order to make reliable
predictions for these amplitudes we should use solutions of the
relativistic three-body problem; at the very least, configuration
mixing of the kind present in the nonrelativistic model must be
included in both the nucleon and all of the final states, as it has
been shown to have substantial effects on the predicted
amplitudes~\cite{KI,CloseLi,Caps}.  It is for this reason that we have
not compared our results directly to the limited available
data~\cite{VBurkert}.

We also note that the Hamiltonian used in Ref.~\cite{CI} is in fact a
three-body mass operator, and so its eigenfunctions can be used
directly and consistently in a calculation of the kind described in
this paper.  We have developed better techniques for performing the
numerical integrals required to obtain the light-front matrix elements
between configuration-mixed states, whose more complicated integrand
structures make standard Monte Carlo methods inefficient.

Nevertheless, it is obvious from the results presented above for
radially excited basis states that electroexcitation amplitudes of the
$P_{11}$ Roper resonance
$N(1440){\textstyle{1\over2}}^+$ and $N(1710){\textstyle{1\over2}}^+$ states,
as well as those of the $\Delta(1600){\textstyle{3\over2}}^+$, are
substantially
modified in a relativistic calculation. Given the controversial nature
of these states~\cite{ZPL,LBL}, we consider this an important result.
Our results show that relativistic effects tend to reduce the
predicted size of the amplitudes for such states at intermediate and
high $Q^2$ values, in keeping with the limited experimental
observations for the best known of these states,
$N(1440){\textstyle{1\over2}}^+$.

Within the context of a model such as this one, full gauge invariance
cannot be achieved.  Reasonable results can be obtained by
constraining one of the higher multipoles in terms of lower
multipoles.  This can be thought of as a variant of the Siegert
hypothesis~\cite{Siegert}, which has been used successfully in the study of
electromagnetic properties of nuclei for many years.

We have also found that the rotational covariance violation is a small fraction
of the larger amplitudes for the $Q^2$ values considered here. In cases where
the dynamics causes an amplitude to be intrinsically small, the uncertainty in
our results for these amplitudes becomes larger. In particular, the calculated
ratios $E2/M1$ and $C2/M1$ for the electroexcitation of the
$\Delta(1232){\textstyle{3\over2}}^+$ in the {\it absence} of configuration
mixing of $D$-wave components into the initial and final state
wavefunctions~\cite{BDW} are probably 100\% uncertain, and are thus consistent
with zero at all $Q^2$~\cite{e2overm1}. This may not be the case in the
presence of such configuration mixing, and we intend to investigate this
possibility, since $\Delta(1232)$ electroproduction is the subject of current
experiments at MIT/Bates and several proposed experiments at
CEBAF~\cite{D1232expt}.

\vskip 12pt

\section*{Acknowledgements}

We would like to thank Mr.~Heng Rao for his help with the analytic
nonrelativistic calculations. We also thank Mr.~Mark Schroeder
for his help with developing message-passing code for parallel
architecture machines.
The calculations reported on here were made possible by grants of
resources from the Pittsburgh Supercomputing Center and the National
Center for Supercomputing Applications.  The work was also partially
supported by the National Science Foundation under grant PHY-9023586 (B.D.K),
by the U.S. Department of Energy through Contract No. DE-AC05-84ER40150 and
Contract No. DE-FG05-86ER40273, and by the Florida State University
Supercomputer Computations Research Institute which is partially funded by
the Department of Energy through Contract No. DE-FC05-85ER250000 (S.C.).

\section*{Appendix A: Conventions for Baryon\hfil\break
 Electroproduction amplitudes}

In the following we have collected the information required to translate
between standard conventions for (pion) electroproduction amplitudes and our
calculated helicity amplitudes.

\subsection*{\bf A.1~~~Pion electroproduction helicity elements}

In the following we relate the helicity elements $A_{l\pm}$, $B_{l\pm}$, and
$C_{l\pm}$ conventionally used to describe pion electroproduction to the
helicity amplitudes defined above in Eq.~(\ref{Alambdas}). The
$A_{l\pm}(B_{l\pm})$ elements are present when a resonance is excited by
photons with their spin projections antiparallel (parallel) to the
target-nucleon spin projection; the momentum of the incoming photon is
parallel to the direction of the initial nucleon spin.
The $C_{l\pm}$ elements are present when a longitudinal photon excites
the resonance (there is no change in the target nucleon spin projection).
These can be written in terms of the helicity amplitudes as
\begin{eqnarray}
A_{l\pm}&=&\mp C^I_{\pi N} f A^N_{\textstyle{1\over2}} \nonumber\\
B_{l\pm}&=&\pm C^I_{\pi N} f \sqrt{16\over (2J-1)(2J+3)}
A^N_{\textstyle{3\over2}}\label{ABC}\\
C_{l\pm}&=&\mp C^I_{\pi N} f C^N_{\textstyle{1\over2}},\nonumber
\end{eqnarray}
where $C^I_{\pi N}$ is an isospin Clebsch-Gordan coefficient for
$N^*\to N \pi$, $J$ is the total angular momentum of the resonance $N^*$,
and $\vert \vec{k}_{\rm c.m.} \vert$ is the c.m. frame three-momentum of
the photon. The $l$ in the subscripts above is the angular momentum of
the relative wavefunction of the final-state $N$ and $\pi$. Note that the
$J$ of the resonance must be $J=l\pm {\textstyle{1\over2}}$, so for a
given $J$ there are two possible $l$ values. However only one of these
values is consistent with parity conservation. For example for the
$\Delta(1232)$ with $J^P={\textstyle{3\over2}}^+$ we have $l=1$ or $l=2$,
but a nucleon (positive intrinsic parity) and a pion (negative intrinsic
parity) in an $l=2$ relative wavefunction have negative overall parity,
so this value is not allowed. The sign in the subscripts above is positive
if $J=l+{\textstyle{1\over2}}$ and negative if $J=l-{\textstyle{1\over2}}$.
So for the $\Delta(1232)$ only $A_{1+}$, $B_{1+}$ and $C_{1+}$ are possible.

The Breit-Wigner factor $f$ for the decay $N^*\to N\pi$ at resonance is
\begin{equation}
f=\left[{1\over (2J+1)\pi}{K_W\over
\vert (\vec{q}_{\pi})_{\rm c.m.}\vert}{M_N\over W}{\Gamma_{N\pi}\over
\Gamma_{\rm tot}^2}\right]^{\textstyle{1\over2}},
\end{equation}
where $(\vec{q}_{\pi})_{\rm c.m.}$ is the pion c.m. frame three-momentum
\begin{equation}
(\vec{q}_{\pi})_{\rm c.m.}^2={[W^2-(m_\pi+M_N)^2][W^2-(m_\pi-M_N)^2]\over
4W^2}.
\end{equation}
With these conventions the $N\gamma$ partial width for a resonance to
decay to nucleon $N$ is
\begin{equation}
\Gamma_{N\gamma}={\vec{k}_{\rm c.m.}^2\over \pi}
{2M_N\over (2J+1)W}\left[\vert A^N_{\textstyle{1\over2}}\vert^2+\vert
A^N_{\textstyle{3\over2}}\vert^2\right].
\end{equation}

\subsection*{\bf A.2~~~Electric, magnetic and scalar multipoles}

In the following we relate the magnetic, electric, and scalar/longitudinal
multipoles $M_{l\pm}$, $E_{l\pm}$, and $S_{l\pm}$ to the helicity elements
described above; these can then be related to the helicity amplitudes
calculated here by use of Eq.~(\ref{ABC}). A transition is defined to be
magnetic with multipoles $M_{l\pm}$ if the total angular momentum absorbed
from the photon is $l$, and electric with multipoles $E_{l\pm}$ if the
total angular momentum absorbed from the photon is $l\pm 1$. The
corresponding scalar/longitudinal multipole is $S_{l\pm}$. If
$J=l+{\textstyle{1\over2}}$ the scattering can be described in terms of
the magnetic, electric and scalar multipole amplitudes $M_{l+}$,
$E_{l+}$, and $S_{l+}$, which can be written in terms of $A_{l+}$,
$B_{l+}$ and $C_{l+}$ as
\begin{eqnarray}
M_{l+}&=&{1\over 2(l+1)}\left[2A_{l+}-(l+2)B_{l+}\right]\nonumber\\
E_{l+}&=&{1\over 2(l+1)}\left[2A_{l+}+lB_{l+}\right]\label{MES+}\\
S_{l+}&=&{1\over l+1}{\vert \vec{k}_{\rm c.m.} \vert\over Q} C_{l+}.\nonumber
\end{eqnarray}
If $J=l-{\textstyle{1\over2}}$ the scattering is described by
$M_{l-}$, $E_{l-}$, and $S_{l-}$ with
\begin{eqnarray}
M_{l-}&=&{1\over 2l}\left[2A_{l-}+(l-1)B_{l-}\right]\nonumber\\
E_{l-}&=&{1\over 2l}\left[-2A_{l-}+(l+1)B_{l-}\right]\label{MES-}\\
S_{l-}&=&-{1\over l}{\vert \vec{k}_{\rm c.m.} \vert\over Q} C_{l-}.\nonumber
\end{eqnarray}
The inverse transformation is
\begin{eqnarray}
A_{l+}&=&{1\over 2}\left[lM_{l+}+(l+2)E_{l+}\right]\nonumber\\
B_{l+}&=&-M_{l+}+E_{l+}\label{MESinv}\\
A_{l-}&=&{1\over 2}\left[(l+1)M_{l-}-(l-1)E_{l-}\right]\nonumber\\
B_{l-}&=&M_{-}+E_{l-}.\nonumber
\end{eqnarray}

\subsection*{\bf A.3~~~Invariant multipoles for $\Delta(1232)$
electroproduction}

For the special case of electroproduction of a $J^P={\textstyle{3\over2}}^+$
particle such as the $\Delta{\textstyle{3\over2}}^+(1232)$, we can define
invariant multipoles $G^*_M$, $G^*_E$
and $G^*_C$. The relation between these and the helicity amplitudes
defined above is
\begin{eqnarray}
G^*_M&=&-F\left[\sqrt{3}A^N_{\textstyle{3\over2}}+
A^N_{\textstyle{1\over2}}\right]\nonumber\\
G^*_E&=&-F\left[{1\over \sqrt{3}}A^N_{\textstyle{3\over2}}-
A^N_{\textstyle{1\over2}}\right]\label{GMEC}\\
G^*_C&=&2{W\over \vert\vec{k}_{\rm c.m.}\vert}F
\sqrt{2}S^N_{\textstyle{1\over2}},\nonumber
\end{eqnarray}
where
\begin{equation}
F=\sqrt{{1\over 4\pi\alpha}{M_N\over W}K_W}
\left[1+{Q^2\over (M_N+W)^2}\right]^{\textstyle{1\over2}}
{M_N\over \vert\vec{k}_{\rm c.m.}\vert}.
\label{F}
\end{equation}
Note that Eq.~(\ref{F}) implies that at the photoproduction point
($Q^2=0$, where $\vert\vec{k}_{\rm c.m.}\vert=K_W$) we have
\begin{equation}
F(0)=\sqrt{1\over 4\pi\alpha}
\sqrt{2M_N^3\over W^2-M_N^2}.
\label{F0}
\end{equation}

\section*{Appendix B: Wavefunctions for the $P$-wave\hfil\break
 baryons}

We construct $N$ and $\Delta$ states which have the correct permutational
symmetry (mixed-$\lambda$ type for $N$ states, and symmetric for
$\Delta$ states) in their flavor/spin/space wavefunctions. A common
totally antisymmetric color wavefunction is assumed, as is an
implicit Clebsch-Gordan sum coupling ${\bf L}$ and ${\bf S}$. The
spin-quartet $N$ states (notation is $\vert F^{2S+1}L_\sigma J^P\rangle$,
with $F$ the flavor $N$ or $\Delta$, and $\sigma$ the spatial
permutational symmetry) $\vert N^4P_M{\textstyle{1\over2}}^-\rangle$,
$\vert N^4P_M{\textstyle{3\over2}}^-\rangle$, and
$\vert N^4P_M{\textstyle{5\over2}}^-\rangle$ are made up of
Clebsch-Gordan linear combinations of
\begin{equation}
uud\chi^S_{{\textstyle{3\over2}},\mu}\Psi^\lambda_{1,M}\ .
\label{N4}
\end{equation}
The spin-doublet $N$ states $\vert N^2P_M{\textstyle{1\over2}}^-\rangle$
and $\vert N^2P_M{\textstyle{3\over2}}^-\rangle$ are made from
\begin{equation}
uud{1\over \sqrt{2}}\left(
\chi^\rho_{{\textstyle{1\over2}},\mu}\Psi^\rho_{1,M}-
\chi^\lambda_{{\textstyle{1\over2}},\mu}\Psi^\lambda_{1,M}
\right),
\label{N2}
\end{equation}
and the spin-doublet $\Delta$ states $\vert
\Delta^2P_M{\textstyle{1\over2}}^-\rangle$ and
$\vert \Delta^2P_M{\textstyle{3\over2}}^-\rangle$ are made from
\begin{equation}
uud{1\over \sqrt{2}}\left(
\chi^\rho_{{\textstyle{1\over2}},\mu}\Psi^\rho_{1,M}+
\chi^\lambda_{{\textstyle{1\over2}},\mu}\Psi^\lambda_{1,M}
\right).
\label{D2}
\end{equation}
Where mixing is allowed, the physical states do not have definite quark
spin $S$. We have used~\cite{IK}
\begin{eqnarray}
\vert N(1535){\textstyle{1\over2}}^- \rangle & \to &
 \phantom{-}0.85 \vert N^2P_M{\textstyle{1\over2}}^-\rangle +0.53
\vert N^4P_M{\textstyle{1\over2}}^-\rangle\nonumber\\
\vert N(1650){\textstyle{1\over2}}^- \rangle & \to &
-0.53 \vert N^2P_M{\textstyle{1\over2}}^-\rangle +0.85
\vert N^4P_M{\textstyle{1\over2}}^-\rangle\nonumber\\
\vert N(1520){\textstyle{3\over2}}^- \rangle & \to &
\phantom{-}0.99 \vert N^2P_M{\textstyle{3\over2}}^-\rangle -0.11
\vert N^4P_M{\textstyle{3\over2}}^-\rangle\\
\vert N(1700){\textstyle{3\over2}}^- \rangle & \to & \phantom{-}0.11
\vert N^2P_M{\textstyle{3\over2}}^-\rangle +0.99
\vert N^4P_M{\textstyle{3\over2}}^-\rangle.
\label{mixing}
\end{eqnarray}

\section*{Appendix C: Nonrelativistic $C^N_{\textstyle{1\over2}}$
amplitudes for\hfil\break
electroexcitation of resonances}

In the Figures above we have plotted predictions of the Isgur-Karl-Koniuk
(nonrelativistic) model for the charge helicity amplitudes
$C^p_{\textstyle{1\over2}}$ and $C^n_{\textstyle{1\over2}}$. To the best of our
knowledge these are not published elsewhere, so we list them below.

For $N{\textstyle{1\over2}}^+(1440)$,
\begin{equation}
C^{(p,n)}_{\textstyle{1\over2}}=A_0\left[-{\sqrt{3}k^2m_q\over
9\alpha_{HO}^2}\right](1,0),
\end{equation}
where
\begin{equation}
A_0=\sqrt{2\pi\over k_0}{e\over 2 m_q}e^{-k^2/(6 \alpha_{HO}^2)},
\end{equation}
and $(k_0, {\bf k})$, with $k=\vert {\bf k}\vert$, is the (virtual)
photon four-momentum in the frame in which the calculation is performed
(taken here to be the Breit frame).
For the states $N(1535){\textstyle{1\over2}}^-$ and
$N(1650){\textstyle{1\over2}}^-$ we give the amplitudes for their two
components $N^2P_M{\textstyle{1\over2}}^-$ and $N^4P_M{\textstyle{1\over2}}^-$,
\begin{eqnarray}
C^{(p,n)}_{\textstyle{1\over2}}
(N^2P_M{\textstyle{1\over2}}^-)&=&A_0\left[i{\sqrt{2}km_q\over
3\alpha_{HO}}\right](1,-1),\\
C^{(p,n)}_{\textstyle{1\over2}}(N^4P_M{\textstyle{1\over2}}^-)&=&0\nonumber,
\end{eqnarray}
and similarly for the states $N(1520){\textstyle{3\over2}}^-$ and
$N(1700){\textstyle{3\over2}}^-$
\begin{eqnarray}
C^{(p,n)}_{\textstyle{1\over2}}(N^2P_M{\textstyle{1\over2}}^-)&=&
A_0\left[i{2km_q\over 3\alpha_{HO}}\right](-1,1),\\
C^{(p,n)}_{\textstyle{1\over2}}(N^4P_M{\textstyle{1\over2}}^-)&=&0\nonumber.
\end{eqnarray}
For $\Delta{\textstyle{1\over2}}^-(1620)$
\begin{equation}
C^N_{\textstyle{1\over2}}=-A_0i{\sqrt{2}km_q\over 3\alpha_{HO}},
\end{equation}
and for $\Delta{\textstyle{3\over2}}^-(1700)$
\begin{equation}
C^N_{\textstyle{1\over2}}=A_0i{2km_q\over 3\alpha_{HO}}.
\end{equation}
For $N{\textstyle{1\over2}}^+(1710)$,
\begin{equation}
C^{(p,n)}_{\textstyle{1\over2}}=A_0\left[{\sqrt{6}k^2m_q\over
18\alpha_{HO}^2}\right](1,-1).
\end{equation}

\vfil\eject


\begin{thebibliography}{99}
\bibitem{e2overm1} Although it may not be obvious from Fig.~\ref{Delta},
our results for the transverse amplitudes at $Q^2=0$ yield E2/M1 and C2/M1
which are {\it exactly} zero.
\end{thebibliography}
\end{document}